\documentclass[preprint]{aastex}
\usepackage{emulateapj5,graphicx,times,color}

\slugcomment{ApJ accepted}

\shorttitle{Extended Ionized Gas Clouds in A1367}
\shortauthors{Yagi et al.}
\begin{document}

\def\Ha{H$\alpha $}
\def\Hb{H$\beta $}
\def\Hd{H$\delta $}
\def\NB{NB}
\def\HaR{NB$-$R}

\def\RAform#1#2#3#4{#1{$^{\rm h}$}#2{$^{\rm m}$}#3.{$^{\rm s}$}#4}
\def\Decform#1#2#3#4{#1$^{\circ}$#2'#3.''#4}
\def\Decform1#1#2#3{#1$^{\circ}$#2'#3''}

\def\NII{[\ion{N}{2}]}
\def\SII{[\ion{S}{2}]}
\def\OI{[\ion{O}{1}]}

\def\HI{\ion{H}{1}}
\def\HII{\ion{H}{2}}
\def\HeI{\ion{He}{1}}

\def\A1367{Abell~1367}

\title{Extended ionized gas clouds in the Abell 1367 cluster
\footnotemark[1]{}}
\footnotetext[1]{Based on data collected at Subaru Telescope, which is operated by the National Astronomical Observatory of Japan.}

\author{
Masafumi Yagi\altaffilmark{2,3}
Michitoshi Yoshida\altaffilmark{4},
Giuseppe Gavazzi\altaffilmark{5},
Yutaka Komiyama\altaffilmark{2,6},
Nobunari Kashikawa\altaffilmark{2,6},
Sadanori Okamura\altaffilmark{3,7},
}

\email{yagi.masafumi@nao.ac.jp}

\altaffiltext{2}
{Optical and Infrared Astronomy Division, 
National Astronomical Observatory of Japan, 
Mitaka, Tokyo, 181-8588, Japan}
\altaffiltext{3}
{Graduate School of Science and Engineering, Hosei University,
3-7-2, Kajinocho, Koganei, Tokyo, 184-8584 Japan}
\altaffiltext{4}
{
Hiroshima Astrophysical Science Center, Hiroshima University,
1-3-1, Kagamiyama, Higashi-Hiroshima, Hiroshima, 739-8526, Japan
}
\altaffiltext{5}{Universit\`a degli Studi di Milano-Bicocca, Piazza della 
Scienza 3, 20126 Milano, Italy}
\altaffiltext{6}
{SOKENDAI (The Graduate University for Advanced Studies), Mitaka,
Tokyo 181-8588, Japan}
\altaffiltext{7}
{Department of Advanced Sciences, Hosei University,
3-7-2, Kajinocho, Koganei, Tokyo, 184-8584 Japan}

\begin{abstract}
We surveyed a central 0.6 deg$^2$ region of Abell~1367 cluster 
for extended ionized gas clouds (EIGs) using 
the Subaru prime-focus camera (Suprime-Cam) with 
a narrow-band filter that covers \Ha.
We discovered six new EIGs in addition to five known EIGs.
We also found that the \Ha\ tail from the blue infalling group (BIG)
is extended to about 330 kpc in projected distance, 
which is about twice longer than previously reported.
Candidates of star-forming 
blobs in the tail are detected.
The properties of the EIG parent galaxies in Abell~1367 
basically resemble those in the Coma cluster.
A noticeable difference is that the number of detached EIGs is
significantly fewer in Abell~1367, while the fraction of 
blue member galaxies is higher.
The results suggest a difference in the evolutionary
stage of the clusters; Abell~1367 is at an earlier stage than the Coma cluster.
\end{abstract}

\keywords{galaxies: evolution --  galaxies: structure --
galaxies: clusters: individual (Abell 1367)
}

\section{Introduction}

Gas loss from star-forming galaxies is 
an important processe in galaxy evolution.
In clusters of galaxies quiescent galaxies outnumber
star-forming galaxies, and 
some gas-loss mechanisms should have worked efficiently.
Various physical mechanisms for the gas-loss have been investigated:
internal consumption by a star formation 
\citep{Larson1980},
ram-pressure stripping \citep{Gunn1972},
viscous stripping \citep{Nulsen1982},
a tidal interaction \citep{Toomre1972,Icke1985,Moore1996,Moore1999},
expelling by a galactic wind caused by
an active galactic nucleus (AGN) and/or starburst \citep{Veilleux2005}, 
etc \citep[see][for review]{Boselli2006}.
If the interstellar gas was not consumed in a galaxy but expelled,
the gas could be found around the galaxy.
Such 
gas around galaxies in cluster
has been found in
X-ray \citep[e.g.,][]{Fabian2003,
Wang2004,Sun2005,Sun2006,Randall2008,Sun2010,Ehlert2013,Gu2013,
Zhang2013,Schellenberger2015},
\Ha\ line \citep[e.g.,][]{
Kenney1995,Gavazzi1995,Kenney1999,
Gavazzi2001,Conselice2001,Yoshida2002,Chemin2005,Cortese2006,
Yagi2007, Sun2007,Kenney2008,Yagi2010,
Sun2010,ArrigoniBattaia2012,Fossati2012,
Yagi2013b,Kenney2014,Yagi2015a,Boselli2016},
radio continuum \citep[e.g.,][]{Gavazzi1978,Miley1980,Kotanyi1983,Gavazzi1984,
Dickey1984,Hummel1991},
\HI\ \citep[e.g.,][]{Oosterloo2005,Hota2007,Chung2007,Scott2013},
and molecular lines \citep[e.g.,][]{Boselli1994,Salome2006,Salome2011,Scott2013,Jachym2014,Scott2015}.
Different phases of the gas are observed at
different wavelengths, while their spatial distributions
often resemble each other \citep[e.g.][]{Sun2010}.

By imaging in \Ha\ narrow-band, we can observe an
ionized gas with high spatial resolution and high sensitivity.
The \Ha\ gas around a galaxy in a cluster 
suggests a recent or ongoing gas-loss event of the parent galaxy,
since the gas is heated by the ambient hot plasma 
and will eventually get mixed with it.
Such extended ionized gas (EIG) was reported 
in various nearby clusters;
in \A1367 \citep[e.g.,][]{Gavazzi2001,IglesiasParamo2002,Sakai2002,Gavazzi2003,Cortese2006,Sakai2012},
in the Virgo cluster \citep[e.g.,][]{Kenney1999,Yoshida2002,Kenney2008,ArrigoniBattaia2012,
Kenney2014},
in the Coma cluster \citep[e.g.,][]{Yagi2007,Yoshida2008,Yagi2010,Yoshida2012,Fossati2012},
in Abell~3627 \citep[e.g.,][]{Sun2007,Sun2010},
and in some distant clusters (e.g.. Abell~851 \citealp{Yagi2015a}).
\citet{Yagi2010} 
found 14 EIGs in the central 1 Mpc of 
the Coma cluster, a nearby rich cluster.
In addition, \citet{Yagi2015a} searched for EIGs in two clusters at z$\sim$0.4.
In one of them, nine EIGs were found, and eight of their parents are 
spectroscopically confirmed members of the cluster.
In the other cluster, no EIG was found.
These studies suggest that 
the relative EIGs number normalized by the cluster richness 
is significantly different.
It is therefore intriguing to investigate 
what characteristics of cluster affect the number of EIGs,
and a survey for EIGs in clusters with different characteristics 
from the Coma cluster is important.

The \A1367 cluster at z = 0.0217 \citep[6494 km s$^{-1}$][]{Gavazzi2010}
lies at the intersection of two filaments \citep{West2000}.
\A1367 is thought to be a young cluster
from its high fraction of spiral galaxies \citep{Butcher1984},
low central galaxy density \citep{Butcher1984},
and the irregular shape of the hot gas distribution
\citep{Jones1979,Bechtold1983,Grebenev1995}.
Moreover, \A1367 would have experienced a multiple merger
of substructures.
From the analysis of 273 redshift measurements, 
\citet{Cortese2004} confirmed that the cluster
has two main density peaks associated with two substructures. 
The northwest subcluster is probably in the early phase of merging into
the southeast substructure.  
\citet{Donnelly1998} presented
the existence of a strong localized shock in 
the intra-cluster medium (ICM).
Chandra observations \citep{Sun2002} 
indicate the presence of cool gas streaming into the cluster core,
supporting a multiple merger scenario.  
The evidence for infall of galaxies at high velocity in the NW subcluster
is provided by three further galaxies: CGCG~097-073, 097-079 and 097-087
that display extended cometary emission exceeding the galaxy length by
approximately 50 kpc in the direction opposite to the cluster center.
\citet{Gavazzi1978} discovered extended radio continuum (1420 MHz) 
emission trailing behind them, further analyzed by \citet{Gavazzi1984},
\citet{Gavazzi1987b}, and \citet{Gavazzi1995}. 
These three galaxies appear marginally \HI\ deficient 
\citep{Gavazzi1987a, Gavazzi1989} and have normal CO
content \citep{Boselli1994}.

Moreover, \A1367 is one of the best-studied nearby clusters in \Ha.
\citet{Gavazzi2001} discovered \Ha\ cometary trails coinciding 
in length and direction with the radio continuum ones.
\citet{IglesiasParamo2002} gives the \Ha\ emitter catalog of 
the \A1367 (including the Coma cluster)
using four exposures of 0.5 degree$^2$ each
obtained with the WFC camera attached to the 2.5m William Herschel
Telescope (WHT), with a sensitivity of approximately 
$10^{-15.5}$ erg s$^{-1}$ cm$^{-2}$ obtained with 
one hour exposures per field. 
The total number of galaxies with \Ha\ in this sensitivity was 41.
\citet{Sakai2002}
discovered a group of dwarf star-forming galaxies near the X-ray
center of \A1367 from  a deep \Ha\ survey, 
and discussed their tidal origin.
\citet{Gavazzi2003} 
independently investigated this star-forming compact group and
suggested that it is in fact infalling onto \A1367 (the group was
dubbed BIG from Blue Infalling Group).  
\citet{Cortese2006} provided
deeper \Ha\ imaging and multi-slit spectroscopy of BIG providing
further evidence for a high velocity encounter between the compact
group and the cluster as a whole.

Thus, many intensive studies in \Ha\ have been performed on \A1367
including a uniform survey in \Ha\ \citep[e.g.,]{IglesiasParamo2002}.
Meanwhile
no uniform survey of EIGs has been carried out yet,
as it requires a deep and wide-field \Ha\ observation.
In this paper, we present a catalog of EIGs 
in the \A1367 cluster found in our \Ha\ narrowband imaging 
with the Subaru Prime Focus Camera (Suprime-Cam).
We assume that the distance modulus to \A1367 is $(m-M)_0 = 35.00$ and 
($h_0$,$\Omega_M$,$\Omega_\lambda$) = (0.697,0.282,0.718)
\citep{Hinshaw2013}.
Under these assumptions, 1 arcsec corresponds to 0.463 kpc.
We use AB-magnitude system of the instrument unless otherwise noted.

\section{Data}

\subsection{Observation}

\begin{figure}[ht]
\includegraphics[scale=0.5,bb=0 0 940 938]{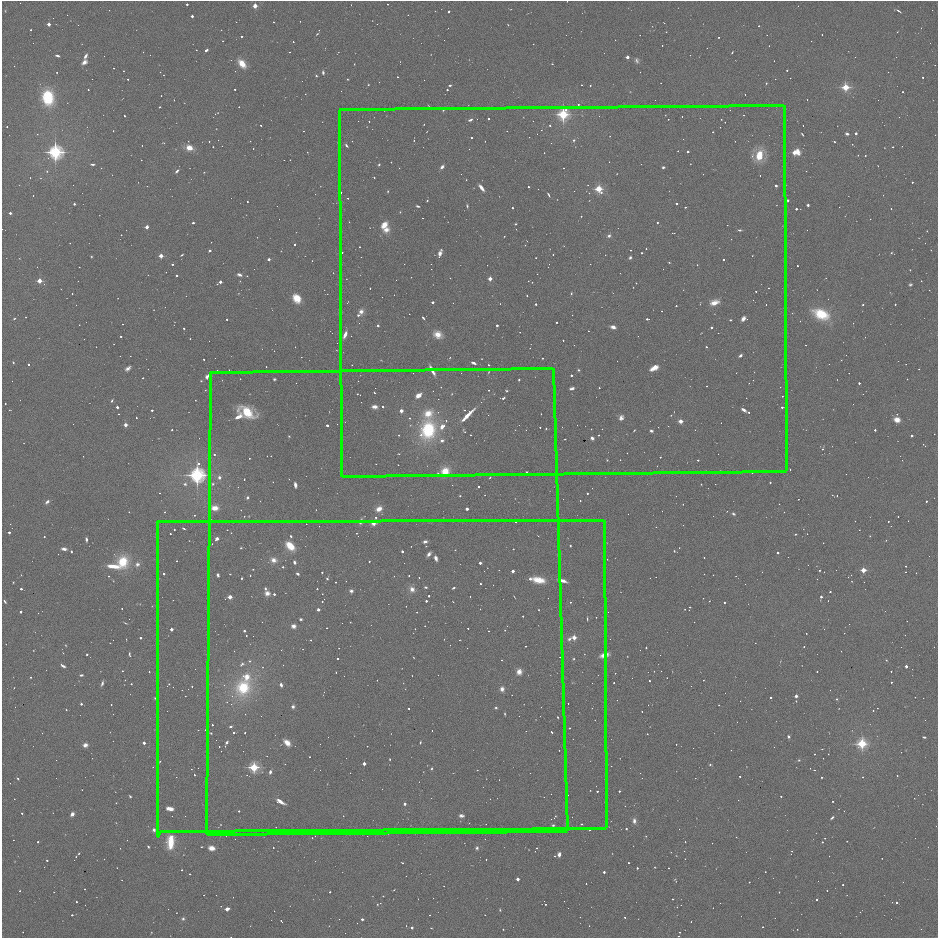}
\caption{\A1367 cluster.
The background image is R-band of Palomar Digitized Sky Survey 2 (DSS2).
The size of the image is 75 arcmin square.
The solid boxes represent the observed regions in this study.
}
\label{fig:FC0}
\end{figure}

Three fields of \A1367 were observed
with Suprime-Cam \citep{Miyazaki2002}
at the Subaru Telescope \citep{Iye2004} in Apr--May 2014 in UTC.
The field position and orientation are shown in Figure \ref{fig:FC0},
and the observations are summarized in Table \ref{tab:obs}.
The pixel scale of the data is 0''.202 pixel$^{-1}$.
The total survey area is 2207 arcmin$^2$,
which corresponds to 1.7 Mpc$^2$.
We dithered to fill the gap between CCDs by 1.1 arcmin, 
typically by five exposures.

We used three broadband filters (B, R, i), and a 
narrowband filter (N-A-L671, hereafter \NB).
The NB-filter was originally designed for observing \Ha\ emitting
objects in the Coma cluster at z = 0.0225,
and has bell-shaped transmittance with a central wavelength of
6712 \AA\ and FWHM of 120 \AA\, \citep{Yagi2007,Yoshida2008}.
Since the redshift of \A1367 (z = 0.0217) is comparable to 
the Coma cluster, the filter is also capable of observing
\Ha\ in \A1367.

\subsection{Data Reduction}

The data reduction was mostly the same as 
\citet{Yagi2010}:
overscan subtraction, flat-fielding, distortion correction,
background subtraction and mosaicking were performed.
The sky background was subtracted with
a mesh size of 512 pixels (1.7 arcmin) square.
We applied several improved processing, too.
Crosstalk is corrected by the method by \citet{Yagi2012},
blooming was masked automatically, 
and some optical ghosts were corrected 
by new algorithms \citep{Yagi2015b}.
For the reference of the bright stars in the field,
PPM-Extended (PPMX) catalog \citep{Roeser2008} was used to obtain the 
position at the epoch of the observation,
after world coordinate system (WCS) 
calibration using astrometry.net \citep{Lang2010}.
The star positions were used for the ghost correction.

\subsection{Photometric Calibration}
\label{sec:photocal}

The flux zero point was calibrated using 
Sloan Digital Sky Survey III (SDSS-III) Ninth Data Release (DR9) 
photometric catalog\citep{DR9}.
The color conversion procedure was the same as 
\citet{Yagi2013a}, and the color conversion coefficients are 
given in \citet{Yoshida2016}.
About 300--900 stars were used for the calibration, and 
the root mean square(rms) was 0.04--0.05 mag.

We measured the flux in a 2'' apertures at $10^6$ random position 
in the combined images and estimated the rms from the 
median of the absolute deviation (MAD) as
rms = 1.4826 $\times$ MAD.
In the combined images, northeast and southwest regions
have no data, and we abandoned the apertures in the regions.
As a result, valid aperture positions were $\sim 7\times 10^5$.
The calculated rms is given in Table \ref{tab:obs}
as a limiting surface brightness (SB).

We adopted the Galactic extinction values of 
0.084, 0.050, 0.050, and 0.039, for B, R, \NB, and i-bands, respectively,
from NASA/IPAC Extragalactic Database (NED)
\footnote{\url{http://ned.ipac.caltech.edu/}},
which uses \citet{Schlafly2011} recalibration of 
\citet{Schlegel1998} data.
The same extinction correction was applied to the whole field.

\subsection{Search for EIGs}
For detecting \Ha\ clouds,
we subtracted the R-band image from the \NB\ image.
Here, we refer to the R-band subtracted \NB-band image as 
\NB-R image, or simply \Ha\ image when it is not confusing.
Note that this \NB-R image includes \NII\ emission,
and possible residual flux of the continuum.
Following \citet{Yagi2010}, we adopted $R-\NB = 0.065$ 
for the R-band subtraction from the \NB\ image.
The color determined the relative flux scaling between $R$ and $\NB$
for a typical continuum spectrum of objects 
without \Ha\ emission/absorption at z$\sim$0.02.

If the underlying continuum is bluer, the $R-\NB$ is smaller
\citep[Figure 3 of ][]{Yagi2010}.
An excess recognized in \NB-R image is thus not always 
an emission but could be a residual of a red continuum.
The \Ha\ flux estimation 
is largely affected by the continuum subtraction.
In this study, we therefore focused on the excess in the 
\NB-R image where continuum is negligible.
Such an excess would be the ionized gas out of galaxies, EIG.
Meanwhile, quantitative discussion about star formation in 
the galactic disk, for example, is beyond the scope of this paper.
Later we investigate possible star formations in EIGs, but
we should be careful that a future spectroscopic confirmation is necessary
\citep[e.g.,][]{Yagi2013b}.

The seeing sizes were estimated using SExtractor\citep{Bertin1996}
version 2.19.5. 
The \NB\ image has a seeing size variation among the field;
0''.7 at NW, while 0''.9 at SE.
The seeing size of most \NB\ data is smaller 
than that of R-band images.
For detecting extended \Ha\ clouds out of galaxies
the difference of the seeing size affects little.
The oversubtraction of R-band is small around pure \Ha\ emissions,
while it is large around objects with continuum (e.g., foreground stars).

In the \Ha\ (\NB-R) image, extended \Ha\ clouds were
searched by visual inspection.
Then B, R, and i-band images were used to reject
possible residuals of bright continuum.
In Table \ref{tab:clouds}, the newly detected \Ha\ emitting clouds
extending beyond galaxies are listed.
Already known EIGs (CGCG~097-073, CGCG~097-079,
CGCG~097-087, CGCG~097-087N, and BIG)
are also shown as a reference.

In the first inspection, 
we did not match the point spread function (PSF),
because the \NB\ images have
better resolution than the R-band images
and PSF matching would make the fine \Ha\ structures blurred.
A slight debris of background subtraction sometimes mimics
an \Ha\ emitting object, but the high spatial resolution of the \Ha\ image
enables us to distinguish artifacts from EIGs.
Moreover, a blur of R-band due to the PSF mismatch 
does not affect the visual inspection 
since we focused on the objects with a weak continuum flux.
In our previous study of the Coma cluster \citep{Yagi2010},
all the EIGs detected in the same procedure as in this study
are spectroscopically confirmed to be real EIGs 
in the Coma cluster (Yoshida et al., in preparation).

As the exposure time and the transparency of the atmosphere
were not uniform in our data, 
the S/N should vary among the positions by $\sim$ 10 percent.
In this study, however, the detection of extended \Ha\ clouds was not 
affected by the S/N variation.
As a test, we made a low S/N image, half of the original, 
by adding artificial noise, and searched for EIGs again.
We can find the same set of the EIGs in the lower S/N image,
though the detail structures of EIGs were buried with the noises
in the eye inspection.
Because we adopted a threshold of a constant SB
in the net \Ha\ and focused on extended features,
the measurement of the EIGs would not largely 
affected by the change of S/N, either.
For another test, we performed a visual inspection
in PSF matched images and confirmed that the detected objects were unchanged.
In the following analyses and figures, 
the PSF matched images were used unless otherwise noted.

\subsection{Conversion of \NB-R to \Ha}
\label{sec:nb2Ha}

Pixel values (renormalized counts)
in \NB-R image are proportional to the \Ha\ SB. 
We calculated pixel value that corresponds to \Ha\ SB of 
$10^{-16}$ s$^{-1}$ cm$^{-2}$ pixel$^{-1}$ 
and converted the \NB-R counts to the \Ha\ SB.
As the \NB\ filter has a bell-shaped transmittance curve,
the response to \Ha\ emission is dependent on the redshift.
In order to take the redshift effect into account, 
we construct a simple model spectral energy distribution (SED) 
for each cloud assuming its redshift
and measure the model magnitude with the \NB\ and R-band filters.
The filter model includes CCD's quantum efficiency,
transmittance of the optics, and the atmospheric transmittance 
\citep{Yagi2013a}.

\begin{figure}[ht]
\includegraphics[scale=0.55,bb=0 0 574 574]{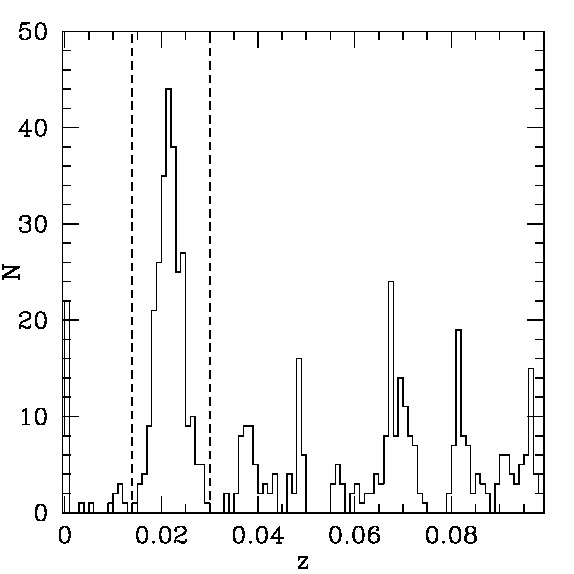}
\vspace{3mm}
\caption{Redshift histogram around \A1367.
The data are taken from SDSS DR9
within 1.5 degree from 
(RA,Dec)(J2000) = (11:44:36.5,+19:45:32).
The vertical broken lines show the redshift range
of \A1367 membership we adopted.
}
\label{fig:Nz}
\end{figure}

Figure \ref{fig:Nz} shows the redshift distribution of 
galaxies taken from SDSS DR12 spectroscopic catalog \citep{DR12}.
The galaxies brighter than $r = 17.8$ mag and within
1.5 degrees from the cluster center are counted.
We adopted RA(J2000) = \RAform{11}{44}{36}{5}, Dec(J2000)=\Decform1{+19}{45}{32}
for the cluster center \citep{Piffaretti2011}.
Based on Figure \ref{fig:Nz},
we defined the galaxies with $0.014<$z$<0.030$ as
the member galaxies of \A1367.
From the distribution, standard deviation of the 
recession velocity is estimated as 815 km s$^{-1}$.
The maximum change of the transmittance of the \NB\ filter
for \Ha\ line is about 0.6 mag (factor $\sim$ 0.6) 
in $0.014<z<0.030$.
The result of the \Ha\ SB and flux in the following analyses 
may suffer $\sim$ 40\% error if the redshift of the emission 
is uncertain.

\begin{figure}[ht]
\includegraphics[scale=0.8,bb=0 0 574 574]{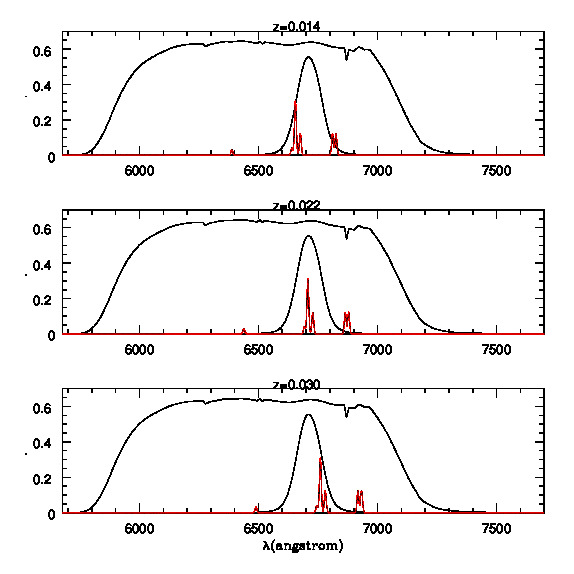}
\vspace{3mm}
\caption{
Redshift dependence of model SED of emission line object 
and the transmittance of NB and R-bands.
The black solid lines show NB and R-band total transmission 
with telescope, optics, and atmospheric transmittance.
The red solid line is our model SED of an EIG after 
continuum subtraction. The flux scale is arbitrary.
The three panels show z=0.014, 0.022 and 0.030 
from the top to the bottom, respectively
}
\label{fig:trans_schematic}
\end{figure}

Spectra of EIGs in previous studies \citep{Yoshida2004,Yagi2007,Yoshida2012}
show wide variety in \NII/\Ha\ ($-1.1<\log($\NII/\Ha$)<0.0$).
The \NII/\Ha\ around BIG given by \citet{Cortese2006} 
also varies within $0.09<$ \NII/\Ha $<0.42$.
In this study, we adopted log(\NII/\Ha) = $-0.4$, i.e. \NII/\Ha = 0.4,
for model SED for the conversion.
The systematic error by this \NII/\Ha\ uncertainty is $\lesssim$60\%.
Other emission lines in R-band, \SII\ and \OI\ are
added in the R-band model magnitude
assuming \SII/\Ha = $-0.4$ and log(\OI/\Ha) = $-1.0$,
according to \citet{Yoshida2012}.
Possible contamination of \SII\ in the \NB-band 
in lower redshift is also taken into account.
The transmittance and the model SEDs at different redshifts
are shown in Figure \ref{fig:trans_schematic}.

\begin{figure}[ht]
\includegraphics[scale=0.8,bb=0 0 574 574]{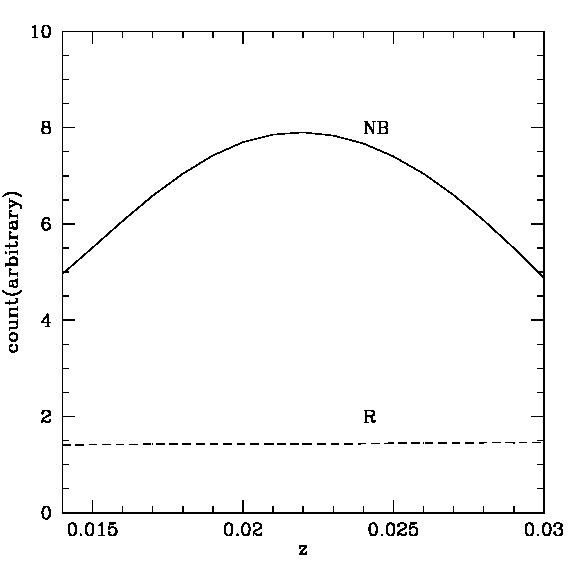}
\vspace{3mm}
\caption{
Redshift dependence of pixel values 
of a model emission line object in NB and R-band data.
The count scale is arbitrary, since the flux zero point 
of the data is changeable. 
Meanwhile, the R-band is scaled to NB-band so that the
flux zero points of the data satisfy R-NB=0.065.
}
\label{fig:zcount}
\end{figure}

We also estimated the oversubtraction of the \Ha\ flux in the R-band 
using the model SED at each redshift.
Because our continuum subtraction is not performed in magnitude 
but in the renormalized counts, the oversubtracted flux is proportional 
to the \NB-R\ pixel value regardless of the \Ha\ equivalent width.
In the redshift range, the oversubtraction is 18--30\%.
It is comparable to the estimation in a z = 0.023 galaxy 
by \citet{Yoshida2016} (23\%).
The redshift dependence is shown as Figure \ref{fig:zcount}.
As expected from Figure \ref{fig:trans_schematic},
the pixel count in R-band is almost constant, while 
that in NB-band changes according to the redshift.
The oversubtraction is corrected 
when converting \NB-R\ counts to the \Ha\ SB.

\subsection{Error of \Ha\ flux}
\label{sec:error}
We estimated the error of the \Ha\ flux of EIGs in two steps.
One is the uncertainty in \NB-R count with
a calibrated zero point, and the other is
the uncertainty in the conversion 
of the \NB-R count to the \Ha\ flux.

The limiting surface brightness is shown in Table \ref{tab:obs},
which is a 1-$\sigma$ fluctuation of the SB
measured in an aperture of 2 arcsec diameter.
The value includes the photon noise of the background sky,
the readout noise, and errors in the data reduction.
The effect of possible remnant of optical ghost 
is also included as an average value.
As the adopted isophote in this study was 
comparable to or larger than the 1-$\sigma$ fluctuation
and the area of the object is much larger than that of 2 arcsec circle,
the S/N of the isophotal flux is high. 
Even for the faintest object, the estimated relative error of 
the count is three percent.

The absolute calibration of \NB\ and R-bands relies on
the SDSS photometry and the color conversion models.
The conversion may have $\sim$ 0.04 mag error in each band
\citep{Yagi2013a}.
The residual of continuum of overlapping objects
(e.g., disk of the parent galaxy) may remain in the \NB-R counts.
In \citet{Yagi2010}, the change of R-NB magnitude of continuum 
is about 0.15 mag peak-to-peak. It means that the error of 
\NB-R count could be $\sim$ 10\% of the rescaled R-band count.
In the regions where the rescaled R-band count is comparable to or 
larger than the \NB-R counts, this error is the dominant one.
We summed the rescaled R-band count where the count is larger than
\NB-R counts, and used 10\% of the sum as 
the estimated error of the \NB-R counts.
The errors given in Table \ref{tab:clouds} 
include the errors described above; the background variation,
the zero-point uncertainty, and the possible continuum error,
assuming that the three errors are independent.

In the conversion of the \NB-R count to the \Ha\ flux,
various uncertainties exist,
since we don't have spectroscopic information of EIGs.
As shown in Figure \ref{fig:trans_schematic},
the redshift should change the observed count in \NB-band.
Thus the error of the redshift can introduce an error
in \Ha\ flux.
As mentioned in Section \ref{sec:nb2Ha},
the error is $\sim$ 40\% if no parent galaxy is identified.
Even if the redshift of the parent galaxy is known,
the EIG may have a recession velocity different 
by several hundred km s$^{-1}$ from the parent
as in the Coma EIGs \citep{Yoshida2012}.
If the offset is 300 km s$^{-1}$, for example,
it may introduce $\gtrsim$ 10\% error.
In the conversion from \NB-R to the \Ha\ flux,
we adopted a model SED with several assumptions in emission line ratios;
\NII/\Ha, \SII/\Ha, and \OI/\Ha.
The uncertainty of \NII/\Ha\ directly affects the conversion,
which may introduce $\sim$ 40\% error.
The uncertainty of \SII/\Ha, and \OI/\Ha\ only affects
the correction of the oversubtraction, which would be $\sim$ 10\%
at most.

\section{Newly Detected Extended Ionized Gas clouds (EIGs)} 
\label{sec:EIGs}

In \A1367, several EIGs were reported in the literature.
CGCG~097-073 and CGCG~097-079 are well known with their prominent tails 
\citep[e.g.,][]{Gavazzi1987b,Gavazzi1989,Boselli1994,Gavazzi1995,Gavazzi1998,Gavazzi2001,Scott2010}.
In \citet{Boselli2014}, we showed a preliminary image 
from the data in this study.
Their prominent tails were well recognized 
in \citet{Gavazzi2001} as a whole.
CGCG~097-087 (UGC~6697) is also the best-studied galaxies in \A1367
\citep[e.g.,][]{Gavazzi1987b,Gavazzi1989,Boselli1994,Gavazzi1995,Gavazzi1998,Gavazzi2001b,Scott2010}. 
\citet{Gavazzi2001b} presented \Ha\ image and spectra of the 
galaxy and a part of the tail with a detailed investigation.
\Ha\ distribution around BIG shows quite complicated morphology,
and has been studied intensively 
\citep[e.g.,][]{IglesiasParamo2002,Sakai2002,Gavazzi2003}. 
\citet{Cortese2006} investigated 
the detailed \Ha\ structure around BIG.

In addition to the well-studied EIGs, we found several new EIGs in \A1367
as given in Table \ref{tab:clouds}.
The distribution of the EIGs is shown in Figure \ref{fig:FC1}.
As most of the EIGs in this study have clear relation to 
a galaxy (parent galaxy) which would be the origin of the gas,
the EIGs are named after their parent galaxy.
The exceptions are the orphan clouds,
and the clouds and the tail around BIG.
The total mass of the ionized clouds is roughly estimated
and given in Table \ref{tab:clouds}. 
The detail of the mass estimation is given in Appendix.

In following subsections,
the new EIGs are shown from the north to the south.
For each EIG, we present 
a B, i, and \Ha\ composite before PSF matching,
and \Ha\ image after PSF matching.
In the \Ha\ image, the green contour represents \Ha\ isophote of 
2.5$\times$ 10$^{-18}$ erg s$^{-1}$ cm$^{-2}$ arcsec$^{-2}$.
The depth is adopted from the previous study in the Coma cluster \citep{Yagi2010}.
To suppress the fluctuation around the isophote, 
we smoothed the image by a Gaussian with $\sigma$ = 2.5 pixels
before the measurement of the isophote.
If the redshift of the parent galaxy was available,
the redshift was used for the calculation of the isophote level.
When it is not available, the cluster's redshift, z = 0.0217, is used.
In order to remove false detections of clump of noises, 
we used SExtractor and only sampled large clumps of \Ha\ excess.
We adopted DETECT\_MINAREA = 500 pixels.
Then remnant of noise clumps and debris of bright star subtraction
are carefully removed by visual inspection of B, R, i and \NB\ images.
The flux and the mean \Ha\ SB are measured inside the contour,
and the extension and the bounding rectangle are measured 
based on the contour.

\subsection{Orphan clouds}

The ``orphan clouds'' 
appear to be floating in the intergalactic space
(Figure \ref{fig:orphans}).

If the clouds are in \A1367, the size of 
the north clump (orphan1) is 33$\times$20 kpc, 
and the south-east clump (orphan2) is 12$\times$3 kpc.
Though the isophote of 2.5$\times$ 10$^{-18}$ erg s$^{-1}$ cm$^{-2}$ arcsec$^{-2}$
at z=0.0217 separates the two clouds, a faint filamentary cloud is 
connecting the clouds and possible star-forming(SF) blobs
are found between the two
as seen in the left panel.
It is uncertain whether they are physically related to each other.

At least within 80 kpc, there are no possible 
parent galaxy candidates.
The nearest giant ($M_r<$-17) galaxy is CGCG~097-102S
whose projected distance is $\sim$ 80 kpc
(Figure \ref{fig:orphan_wide}).
Such parent-less clouds were not found in the Coma cluster
\citep{Yagi2010}, or in Abell~851 \citep{Yagi2015a}.
In \A1367, the tip of EIG of CGCG~097-083 is $\sim$ 170 kpc in 
the projected distance from the center of the parent galaxy.
The tip however seems connected to the parent galaxy as a long tail.

There are several questions about the orphan clouds;
the ionizing source and the origin of the gas.
The ionizing source is a big problem for all EIGs.
Possible mechanisms are a star-formation in situ, 
illumination of the young stars in the parent galaxy,
illumination of AGN, inside shock heating, UV from ambient hot gas,
etc.
Since the orphan clouds are at least 80 kpc away from giant galaxies,
a UV from parent galaxy disk is unlikely.
A possible star formation around the cloud is seen in a part of the clouds.
In Figure \ref{fig:orphans},
\Ha\ emission and blue continuum is recognized near orphan2,
which is marked as SF blobs.

The region is also visible in Galaxy Evolution Explorer (GALEX) 
NUV image in the archive\footnote{\url{http://galex.stsci.edu/GR6/}}.
If the SF blobs are the ionizing source of the whole clouds, however, 
a gradient of \Ha\ SB is expected.
The SB should be bright near the region 
and fainter in distant regions.
No such gradient is seen in the data,
and thus the ionization by the SF blobs is unlikely.
More details of the {SF blobs} are discussed 
in Section \ref{sec:starformation}.

For examining the possibility of AGN illumination,
we compared them with an example of a known AGN illuminated cloud,
Hanny's Voorwerp \citep{Lintott2009,Keel2012} at z=0.050.
Hanny's Voorwerp is a famous 
extended ionizing cloud in a less crowded environment
found in SDSS imaging data by the Galaxy Zoo project \citep{Lintott2008}.
The ionizing source of Hanny's Voorwerp was revealed
to be AGN of the parent galaxy(IC~2497) 50 kpc away.
The size of the Hanny's Voorwerp (18$\times$33 kpc) is comparable to
that of orphan1($\sim 20\times 33$ kpc).
Meanwhile, Hanny's Voorwerp is much brighter than the orphan clouds.
According to Table 2 of \citet{Lintott2009}, 
the mean surface brightness of \Ha+\NII\ 
is $5\times 10^{-16}$ erg s$^{-1}$cm$^{-2}$arcsec$^{-2}$, 
which corresponds to 21.7 mag arcsec$^{-2}$ in our NB-R image.
The mean surface brightness of the orphan clouds is 
26.6 mag arcsec$^{-2}$, about 1/90 in flux.
The nearest AGN from the orphan clouds 
is CGCG~097-121 at $\sim$ 180 kpc in \citet{Gavazzi2011},
which is classified as a LINER.
If the orphan clouds are also ionized by the AGN,
the ionizing flux from the AGN would be weaker by an order or more
than that at Hanny's Voorwerp because of the distance.
Added to that, if the ionizing flux of CGCG~097-121 is an order 
weaker than the source AGN of Hanny's Voorwerp,
the mean surface brightness of $\sim 27$ mag arcsec$^{-2}$ 
of the orphan clouds could be reproduced.
Thus the AGN illumination is one of promising scenarios.
Shock heating and UV from the hot gas are 
also promising ionizing mechanisms.

The origin of the gas is another question.
They would have been ejected from a galaxy,
but the mechanism is unclear.
It can be a ram-pressure stripping, 
a jet, a wind, a tidal interaction, etc.
The clouds are away from their parent at least by 80 kpc,
if their parent is a giant galaxy.
Assuming that their speed to the parent galaxy
is $\lesssim$ 1000 km s$^{-1}$, 
they should have survived in the cluster environment
for $\gtrsim$ 70 Myrs.
As no counterpart is found near the clouds
in \HI\ survey of \A1367 \citep{Scott2010},
it is rather unlikely that the orphan clouds are re-ionized portions 
of a long tail, such as the long tail of NGC~4388 in the 
Virgo cluster \citep{Yoshida2002, Oosterloo2005}.

Another possibility is that the parent galaxy is
a very faint dwarf.
A possible ultra diffuse galaxy 
\citep[UDG; e.g.,][]{vanDokkum2015a,vanDokkum2015b,Koda2015,Yagi2016}
is detected near the possible star-forming region,
and the elongation of the SF blobs seems to align to the galaxy.
The galaxy is visible in Figure \ref{fig:orphans} bottom-left,
but shows no \Ha\ emission.
At \A1367 distance, the R-band absolute magnitude of the galaxy is
$M_R\sim$ -14.5 mag and its effective radius is $\sim$ 4 kpc,
measured by GALFIT \citep{Peng2002,Peng2010}.

If the redshift of the UDG is comparable to the SF blobs,
the SF blobs may be associated to the galaxy, and it will be
a hint to understand the formation of UDGs.
Meanwhile, the shape of the whole orphan clouds is 
unlikely to have been stripped from the possible UDG.
There are many other faint dwarfs whose redshift is unknown.
Current data are insufficient to investigate more on the parent galaxy 
of the clouds.

\subsection{CGCG~097-092}

CGCG~097-092 (Figure \ref{fig:097092}) 
shows a EIG of a cone-like appearance
extended to $\sim$ 30 kpc.
The parent galaxy does not show a sign of AGN
\citep{Gavazzi2011}, while 
the spectrum of the center of the parent galaxy shows 
a starburst feature.
The UV light from the starburst would be 
the source of the ionization.
In Table \ref{tab:clouds}, we also showed the property of 
the tail excluding the core region.
The origin of the EIG could be a ram-pressure stripping 
or a superwind by the central starburst\citep{Veilleux2005}.
The two will be distinguished if the 
recession velocity profile will be available in future.
The velocity of EIG should be comparable to the parent near the
galaxy if the ram-pressure is the mechanism, while 
the difference should be seen if a wind is the origin.

The direction of the tail is somehow ``toward'' the 
cluster center; the angle is about 30 degrees.
The recession velocity of the parent galaxy 
has small difference from that of the cluster
($\sim$ -40 km s$^{-1}$).
If ram-pressure stripping is the origin,
it suggests that the parent galaxy moves away from the 
cluster center in the projected plane.

\subsection{2MASX~J11443212+2006238}

The EIG of 2MASX~J11443212+2006238 was detected
at the very edge of our field \ref{fig:FC1}. 
The EIG may have a much longer extension.
In Figure \ref{fig:2MASXJ},
residuals of CCD chip edge pattern remain.
The sharp cutoff on the left would be an artifact.
Thus the size and total \Ha\ flux are unreliable.
Meanwhile, the S/N of the EIG is enough, and 
the mean SB is reliable.

SDSS spectrum of the parent galaxy
shows clear \Ha\ emission from the core,
and \citet{Gavazzi2011} classified it as HII. 
\citet{Kriwattanawong2011} 
also reported \Ha\ from the galaxy.

Interestingly, the parent galaxy is red,
as shown in Section \ref{sec:CMD}.
We checked $g-i$ color in \citet{Consolandi2016}
and confirmed that $g-i$ is also red ($g-i = 1.03$).
In the \NB-R image, the central $\sim$15'' 
region shows an excess,
which may be the remaining star formation,
though the \NB-R color is unreliable for \Ha\ estimator
in the galactic disk.
In the disk, no obvious spiral pattern is recognized in \Ha.
In the \Ha\ image,
the southern half of the galaxy shows a sign of negative \NB-R,
while positive in the northern half.
This morphology resembles IC~4040 in the Coma cluster \citep{Yoshida2008}
and implies that 2MASX~J11443212+2006238 suffers 
strong ram-pressure stripping from the south direction.
The red color and negative \NB-R may be implying that this galaxy is partly 
in a transition phase from star-forming galaxy to post-starforming galaxy.

\subsection{\Ha\ tail from BIG}

We found that the \Ha\ tail from BIG continued about twice longer 
than reported by \citet{Cortese2006} reaching
a projected distance of $\sim$ 330 kpc (Figure \ref{fig:BIGtail}).
The tail was called ``NW'' in \citet{Cortese2006}.

In Table \ref{tab:clouds}, two measurements are given.
We first measured the whole \Ha\ 
including BIG regions, Then, we set the north-west boundary of 
BIG as the knot named K1 in \citet{Gavazzi2003} and \citet{Cortese2006},
and measured the tail out of the boundary.
Total \Ha\ gas mass of BIG+tail is calculated to be 
$(8\pm 5)\times10^{10} \sqrt{f_v} M_\sun$, 
where $f_v$ is the volume filling factor.
Excluding the \Ha\ around BIG,
the tail mass would be $\sim (3\pm 2)\times10^{9} \sqrt{f_v} M_\sun$.
The long tail implies that the group would have 
suffered the ram-pressure from the intra-cluster gas of \A1367
for several hundred Myrs, 
while the group have moved roughly straight toward the southeast.

The ionizing source of long \Ha\ tail is a big mystery in EIGs 
\citep[e.g.,][]{Yoshida2004,Cortese2006,Yagi2007,Kenney2008}. 
\citet{Boselli2016} discussed that
the ionizing mechanism of 80 kpc long tail of NGC~4569 
in the Virgo cluster could be shocks, MHD waves, or heat conduction.
The discovery of the 330 kpc-long \Ha\ tail of BIG 
sets constraints on possible models.
In B, R, and \NB\ composite image, 
we detected several SF blob candidates.
The zoomed-up images are shown in Figure \ref{fig:BIGSFR}.
All of them have counterparts in GALEX NUV data.
Though they may ionize some gas around,
they are insufficient to ionize the distant part of the tail.
The details of the SF blobs are investigated later in Section 
\ref{sec:starformation}.

As seen in Figure \ref{fig:BIGtail},
the tail of BIG shows a sign of winding
with $\sim$ 40 kpc width.
An idea to explain the morphology
is that the winding results from the internal motion 
of the galaxy in the group.
Though merger and interactions among the members of BIG is
suggested \citep{Cortese2006},
here we calculate a simple model.
Assuming that a typical velocity of group member galaxies 
around the center of the group 
is $\sim$ 200 km s$^{-1}$ \citep{Gavazzi2003},
the crossing time of 40 kpc would be $\sim$ 200 Myr.
In the \Ha\ image, 
one or two possible periods are barely recognized in the tail.
The projected length of 330 kpc tail might 
have been formed in 400--800 Myr,
and the tangential velocity of BIG would be 400--800 km s$^{-1}$.
As a result, the \Ha\ tails form helices,
and we see them as winding shapes by projection.
This speculation can be examined by measuring the
velocity profile of the tail along the winding
by future spectroscopic observations.

\subsection{CGCG~097-093}

CGCG~097-093  (Figure \ref{fig:097093}) shows 
a morphology like jellyfish galaxies
\citep[e.g.,][]{Cortese2007,Smith2010,Ebeling2014,Poggianti2016}.
Enhancement of faint blue stars and \Ha\ emission are 
seen to the northeast.
Inside of the disk, an asymmetry of spiral arms
is recognized, which may be a sign of a recent tidal interaction.
The recession velocity of CGCG~097-093 is 7298 km s$^{-1}$.
Its south-west neighbor, CGCG~097-088 is 5616  km s$^{-1}$,
and north neighbor CGCG~097-094 is 7998 km s$^{-1}$.
If the \Ha\ ejection is affected by interaction with 
the neighbor, it would be with CGCG~097-094.

Meanwhile the morphology in \Ha\ suggests ram-pressure origin, 
as the galaxy-wide gas flow would be difficult to 
be made by a tidal interaction only.
Though the asymmetric stellar distribution suggests 
a tidal interaction, some ram-pressure stripped galaxies 
also show the asymmetry \citep{Abramson2014,Jachym2014}.
\citet{Vollmer2003} suggested 
a mixture of a tidal interaction and a ram pressure 
for NGC~4656 in the Virgo cluster
which shows an extended gas tail and an asymmetric stellar distribution.

\subsection{CGCG~097-122}

CGCG~097-122 (NGC~3859) looks like an edge-on spiral galaxy with 
a distorted faint halo which was revealed by
our deep imaging (Figures \ref{fig:097122}).
The halo is elongated toward the northeast 
almost along the major axis of the galaxy. 
At the southwest side of the galaxy, the halo is extended 
to the projected distance of $\sim$10 kpc above the galaxy disk. 
At the north of the galaxy, a small galaxy 
which is possibly an interacting dwarf companion is seen. 
A part of the extended halo may be formed by the interaction
between this companion and CGCG~097-122.

The bright disk \Ha\ emission together with the
very blue color of the galaxy and the emission line 
dominated nuclear spectrum indicate that 
the star formation of this galaxy is very active. 
Its nuclear spectrum is typical of starburst galaxies 
\citep{Gavazzi2013} 
and has no sign of AGN activity \citep{Gavazzi2011}. 
The total \Ha+\NII flux of CGCG~097-122 is 
$10^{-12.65}$ erg s$^{-1}$ cm$^{-2}$ \citep{Gavazzi2006},
and \NII/\Ha = 0.36 \citep{Gavazzi2013}.
The \Ha\ luminosity is calculated to be 
$\sim 1.9 \times 10^{41}$ erg s$^{-1}$,
which is slightly larger than the typical value of the
\Ha\ luminosity of nearby starburst galaxies
\citep[$\sim 5$--$10 \times 10^{40}$ erg s$^{-1}$; e.g.,][]{Strickland2004}.

The morphology of the \Ha\ emission is peculiar 
(Figures \ref{fig:097122}).
It is characterized by its two narrow spurs 
whose widths are $\sim$2 kpc extended $\sim$10 kpc to the northwest and 
the southeast from the galaxy disk, and strong disk \Ha\ emission. 
The spurs look like two wings of a flying bird. 
The north part of the \Ha\ nebula of CGCG~097-122 seems to be 
flowing toward the northeast.
In addition, the two spurs also are bending to the same direction. 
These morphological characteristics suggest that the
hot gas of the galaxy is pushed by the ram pressure 
from the southwest.
However, in contrast to the extended nebulae seen around
other galaxies in this study,
there is no clear indication of one-sided elongation of the ionized gas. 
There is no sharp cutoff at the southwest edge of the \Ha\ nebula,
and no extended part of \Ha\ out of the galaxy at the northeast,
but rather shows an extension toward east-southeast.
In addition, the northern faint flowing part of the nebular continuum 
is overlapped by the dwarf companion mentioned above suggesting that 
this part of the nebula could be formed by a galaxy-galaxy interaction 
rather than ram pressure. 
Therefore, the ram pressure stripping may not be strong enough 
to create the overall structure of the EIG around CGCG~097-122, 
though it possibly affects the slight bending of the spurs and the positional
offset between the bright part of the \Ha\ and the stellar continuum.

The two narrow spurs would be parts of a bipolar bubble driven by the
nuclear starburst. 
It is well known that active star formation
drives a large-scale outflow from the star-forming region; 
a starburst superwind \citep{Veilleux2005}.
In its early phase, the outflow forms a large bubble of ionized gas, 
a superbubble, then the bubble is broken due to a continuous energy input 
from the starburst \citep[e.g.,][]{Strickland2000}. 
The morphology and size of the spurs resemble 
broken superbubbles observed in nearby starburst galaxies such as 
NGC~253 \citep{Strickland2002}, 
NGC~1482 \citep{Hoopes1999}, 
and NGC~3628 \citep{Fabbiano1990}. 
Sometimes the superbubble or the superwind
has an antisymmetric morphology in the ionized gas emission; 
one side of the bubble is much brighter than the other side. 
For example, the superwind of NGC~3628 is seen as a long plume 
from the galaxy disk \citep{Fabbiano1990}.
The large-scale faint \Ha\ nebula of NGC~253
is characterized by the $\sim$9 kpc long bidirectional plumes 
extending from the disk \citep{Strickland2002}. 
The morphology and size of the two spurs of CGCG~097-122 are 
quite similar to these features, suggesting
that the spurs are originate in a starburst superbubble/superwind.

\section{Results}

We compared the EIGs and their parents with 
those in the Coma cluster \citep{Yagi2010}.
The redshifts of \A1367 and the Coma cluster are 
almost the same; z=0.0217 and 0.0225,
and the two data were obtained with
the filter sets, the instrument and the telescope, 
and analyzed in the same way.
Regarding the data quality,
the 1$\sigma$ of the SB in a 2 arcsec aperture 
in NB and R bands in our previous study in the Coma
were 27.5(NB) and 28.1(R) mag arcsec$^{-2}$\citep{Yagi2010},
while they were 27.7(NB) and 27.9(R) mag arcsec$^{-2}$ in \A1367 in this study, 
respectively.
The difference of S/N in NB-R image is about six percent.
The small difference is negligible compared with other factors.
The projected survey area was 1.2 Mpc$^2$(Coma) and 1.7 Mpc$^2$(\A1367),
respectively.

In this section, we give several results from the comparison
of EIGs and their parents in \A1367 and in the Coma.
Interpretation of the result is given later 
in Section \ref{sec:comparison}.

\subsection{EIG length}

We found that EIGs in \A1367 are typically longer than those in the Coma.
In Figure \ref{fig:length}, the cluster-centric projected distance versus 
the long side of the bounding box of EIGs in both clusters are plotted.
It does not mean that the EIG is continuously extended to the length.
The length rather approximates the distance of the tip of EIG from the parent.
As the virial radius of the Coma is larger than that of \A1367,
the difference gets larger if we normalize it by 
the cluster virial radius.
In Figure \ref{fig:length}, the longest one is the tail of BIG.
The parents of other long ($>$50 kpc) EIGs are 
three bright spirals, CGCG~097-073, 097-079, and 097-087 in \A1367,
and GMP2559(IC4040), GMP4060(RB199) and GMP2910(D100) in the Coma.
All of them are thought to be made by ram-pressure stripping
\citep{Gavazzi2001,Gavazzi2001b,Yagi2007,Yoshida2008,Yoshida2012}.
Among them, two in the Coma, GMP4060 and GMP2910, 
show a post-starburst signature.

\subsection{Color of parent galaxies and EIG parent fraction}
\label{sec:CMD}

For investigating the color of parent galaxies,
we first used SDSS DR12 spectroscopic catalog
for the member selection of \A1367.
Selection criteria were $0.014<$z$<0.030$, $r<17.7$ mag,
and in the region where our observation covered.
We refer to the list as an ``unbiased member list.''
Though the spectroscopic target selection of SDSS is not complete,
it is expected to be unbiased.
Next, the galaxies in SDSS DR12 photometric catalog 
whose spectroscopic information is not 
available in SDSS DR12 were cross-identified 
with other spectroscopic catalogs 
\citep{IglesiasParamo2002,Gavazzi2003,Cortese2003,Cortese2004,Cortese2008,Smith2004,Kriwattanawong2011}, 
and added if their redshift satisfied $0.014<$z$<0.030$
and $r<17.7$ mag.
We used our Suprime-Cam data for visual inspection 
of the cross-identification, and thus 
only the galaxies in our observed field are added.
Though the completeness is higher in the list,
the sampling is heterogeneous.
We refer to the sample as an ``extended member list.''
The color magnitude diagram (CMD) of the member galaxies 
is shown as Figure \ref{fig:CMD} using the extended member list.
The AB magnitude offset adopted by K-correct \citep{Blanton2007} v4;
$m_{AB}-m_{SDSS}$ = 0.012(g) and 0.010(r),
and the Galactic extinction correction
($A_{g}$ = 0.076 and $A_{r}$ = 0.053) are applied.
The color magnitude relation (CMR) of early type members 
of \A1367 is 
\begin{equation}
g-r = 1.210 -0.0304 r
\end{equation}
which is obtained from fitting to the data.
If we classify the galaxies into red/blue at 0.2 mag redder
than the CMR (broken line in Figure \ref{fig:CMD}), 
six parents are blue and two (2MASX~J11443212+2006238 and CGCG~097-122)
are red.
Regarding the BIG members, CGCG~097-114 is blue,
while CGCG~097-120 and CGCG~097-125 are red. 
SDSS~J114501.81+194549.4 is marginally (by 0.01 mag) red.

The number of members and EIGs are given in Table \ref{tab:stat}.
For statistical test, we didn't count CGCG~097-087N,
as it isn't in the unbiased member list.
The CMR and color threshold of the Coma is taken from \citet{Yagi2010},
and the unbiased member list of the Coma is constructed from 
SDSS DR12 applying criteria that $0.015<$z$<0.035$, $r<17.7$ mag,
and in the region where the observation covered.

Because the population size is very small, most of the differences 
are within statistical error. 
Only the blue member fraction of \A1367 is 
significantly higher than that of Coma.
The whole EIG parent fraction is six percent in the Coma and 
nine percent in \A1367. The difference is marginal ($p=0.06$, upper).

\subsection{Distribution in velocity-distance plane}

Figure \ref{fig:vd} shows the distribution of 
galaxies in the projected distance from the cluster center 
versus the recession velocity plane.
The galaxies with a spectroscopic redshift by SDSS are plotted.
Many of the parents have a large difference 
from the cluster's recession velocity.
This trend is similar to that in the Coma cluster \citep{Yagi2010}.

\begin{figure}[ht]
\includegraphics[scale=0.45,bb=0 0 1036 1118]{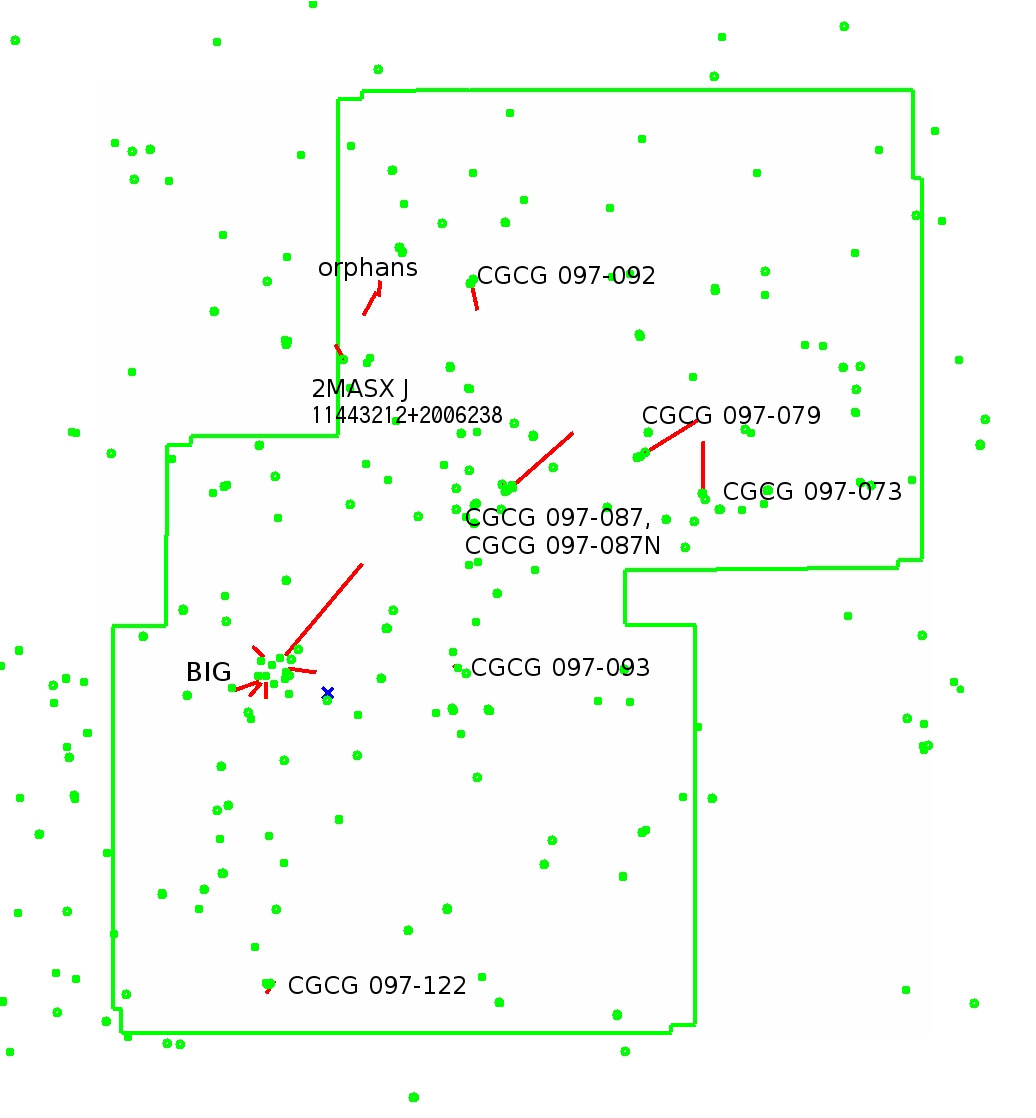}
\caption{Distribution of EIGs.
The green dots are spectroscopic member galaxies.
The green polygon shows the observed region.
The blue x-mark shows the center of the cluster \citep{Piffaretti2011}.
Red lines show the major direction and length of EIGs.
Each parent name is also shown.
The clouds whose parents are uncertain are labeled as 
``orphans.''
}
\label{fig:FC1}
\end{figure}

\begin{figure}[ht]
\includegraphics[scale=4.2,bb=0 0 62 71]{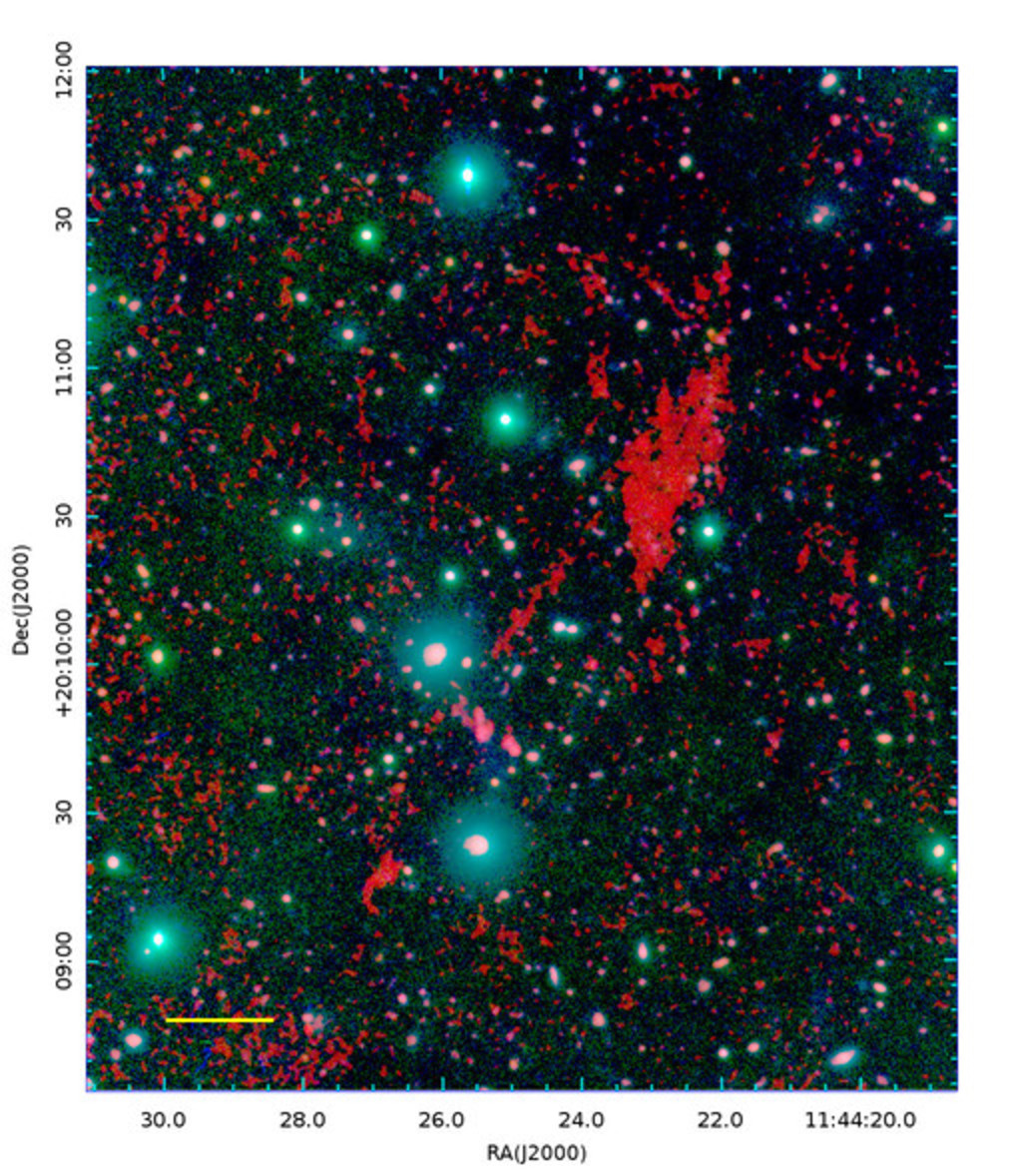}
\includegraphics[scale=4.2,bb=0 0 62 71]{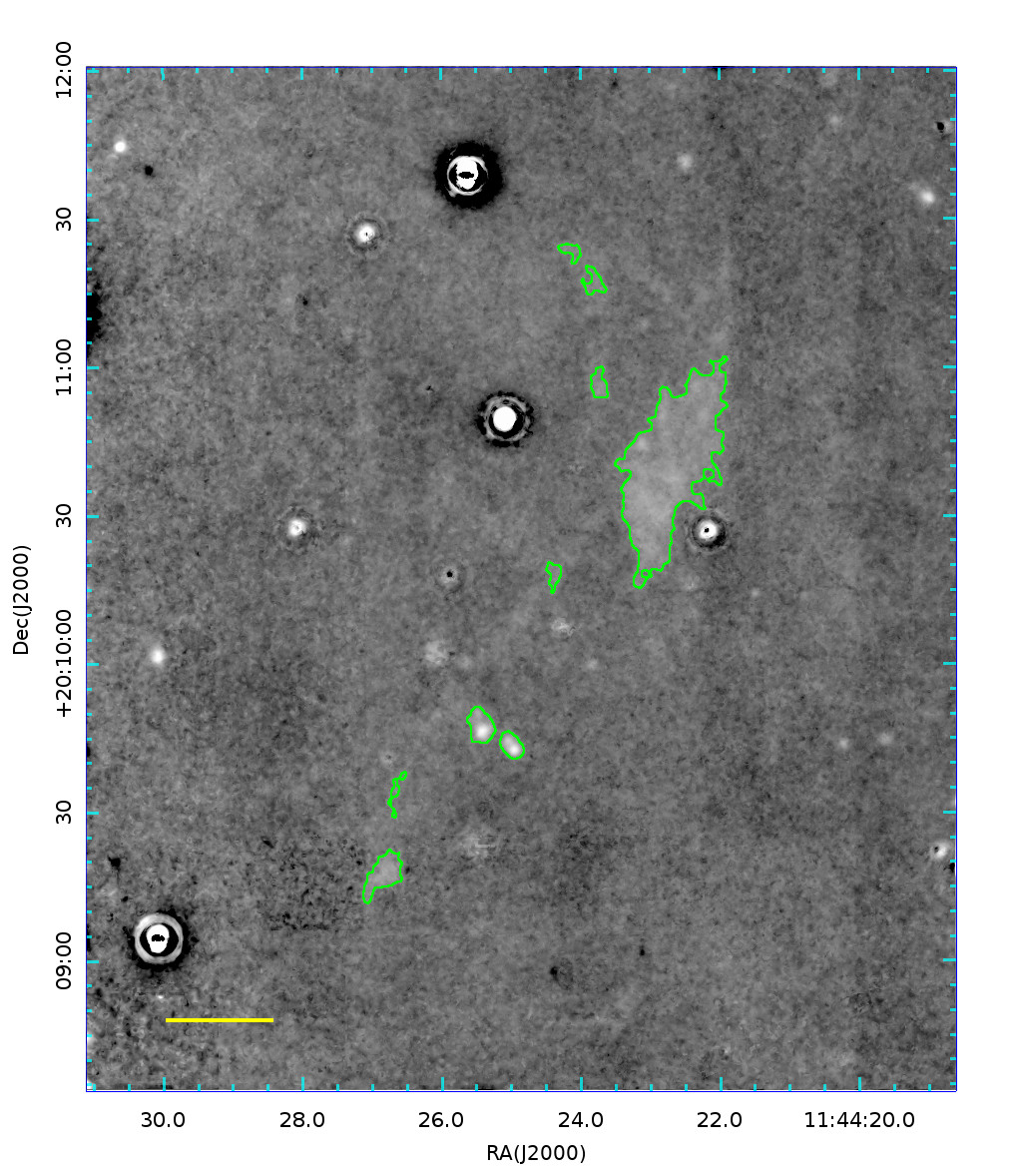}\\
\includegraphics[scale=4.2,bb=0 0 62 71]{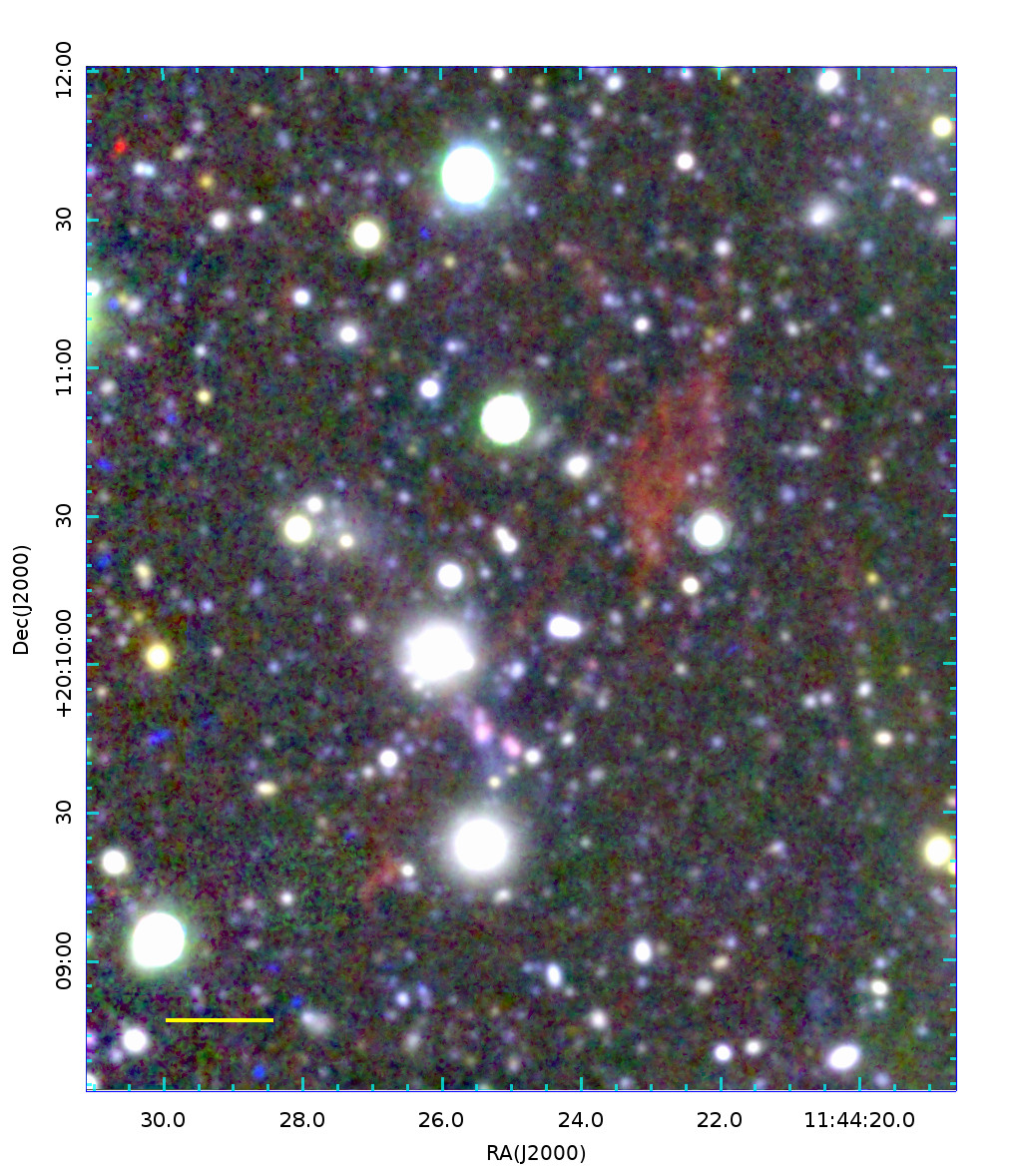}
\includegraphics[scale=4.2,bb=0 0 62 71]{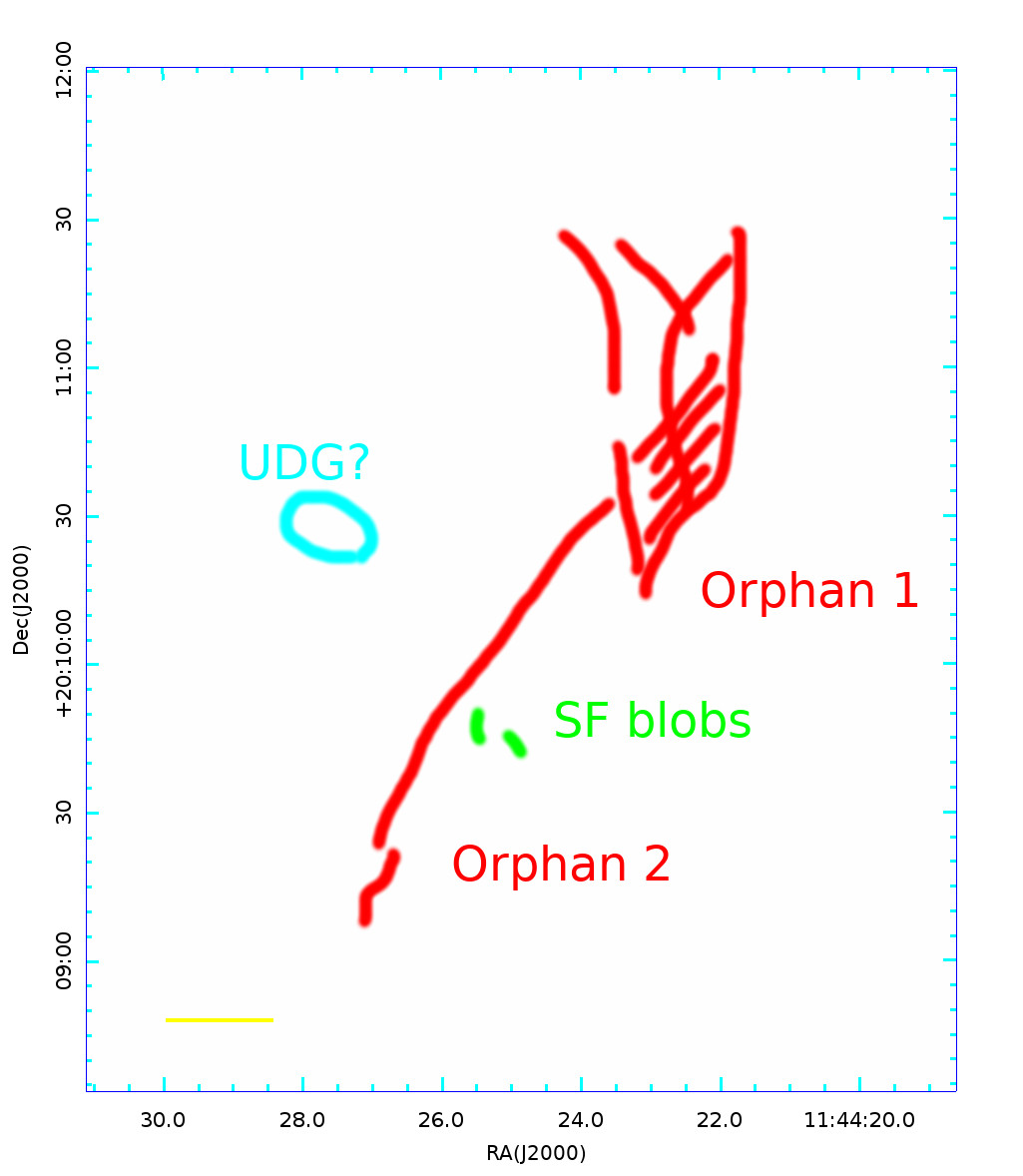}
\caption{ 
(top-left) 
B, i, and \NB-R(\Ha) three-color composite around the orphan clouds.
North is up and east to the left.
PSF is not matched to obtain higher spatial resolution.
Contrast of each color is arbitrary.
The yellow bar at the bottom left shows 10 kpc.
(top-right) 
\Ha(\NB-R) image of the same region after PSF matching.
White indicates \NB\ flux excess and black indicates 
\NB\ flux deficiency.
Typical $z\sim 0.022$ galaxies without \Ha\ emission
have the same color as sky background.
As discussed in the text, 
redder and bluer underlying stellar components 
made \NB\ excess and \NB\ deficiency, respectively.
The green contour represents the isophote of 
2.5$\times 10^{-18}$ erg s$^{-1}$ cm$^{-2}$ arcsec$^{-2}$
assuming z=0.0217.
The PSF matching made the spatial resolution lower 
and the detailed faint feature seen in top-left panel 
is under the threshold.
(bottom-left)
PSF matched B, R, and \NB\ composite.
The relative B, R, and \NB\, scales are set so that 
typical galaxies at z = 0.022 without \Ha\ emission 
($B-R$ = 1.0, and $R-\NB$ = 0.065) are gray.
The magenta objects would be SF blobs.
(bottom-right)
Schematic figure around the orphan clouds.
}
\label{fig:orphans}
\end{figure}

\begin{figure}[ht]
\includegraphics[scale=0.34,bb=0 0 680 674]{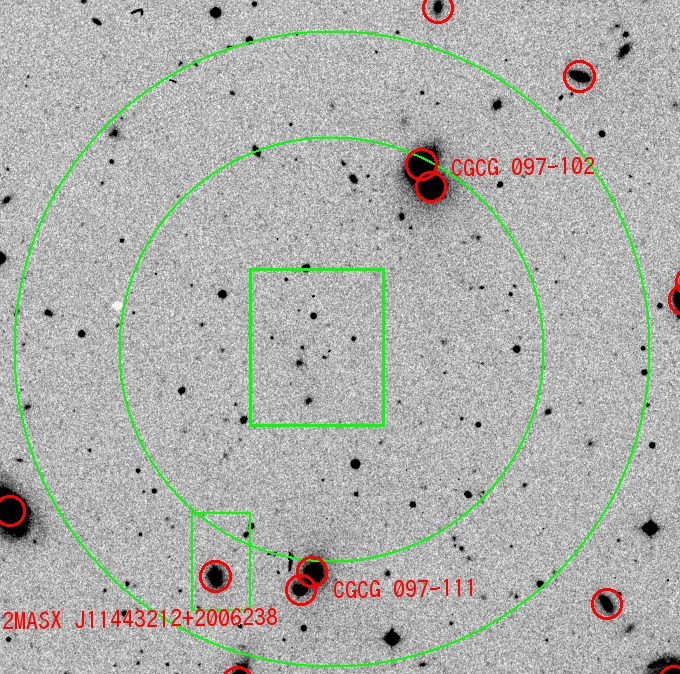}
\includegraphics[scale=0.285,bb=0 0 807 791]{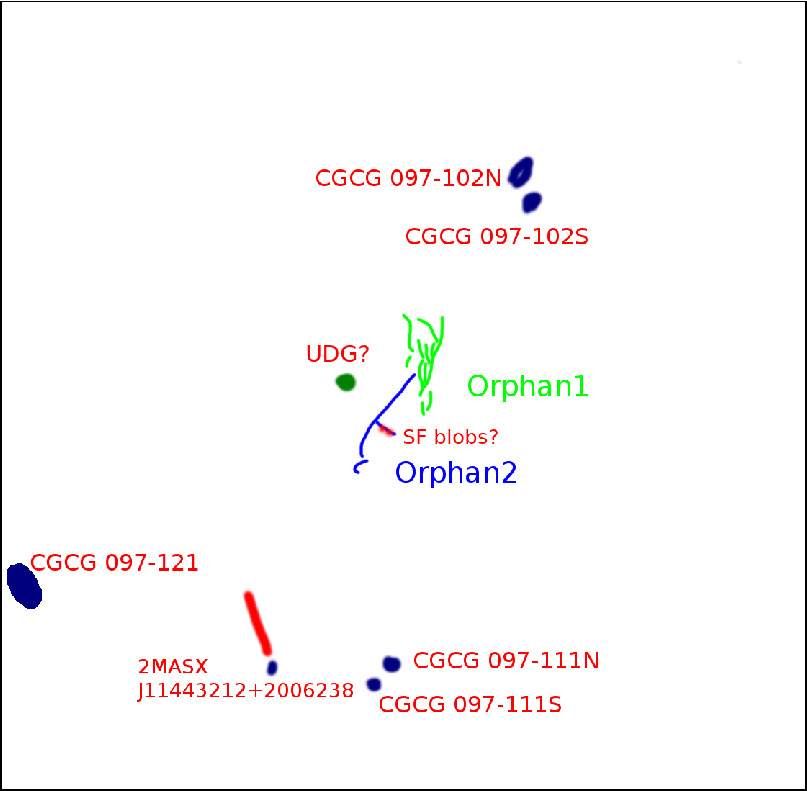}\\
\caption{Member galaxies around the orphan clouds.
(left) DSS2 image. Member galaxies are marked with red circles.
The central rectangle shows the region of Figure \ref{fig:orphans},
and the left-bottom box shows that of Figure \ref{fig:2MASXJ}.
Two circles show scale of 100 kpc and 150 kpc from the orphan clouds.
The names of member galaxies within 150kpc from the clouds are also shown.
(right) Schematic figure around the clouds. The image scale is 
the same as the left panel. 
The ``UDG'' is not visible in top panels
but is seen in Figure \ref{fig:orphans}.
}
\label{fig:orphan_wide}
\end{figure}

The two parents near the cluster's velocity are 
CGCG~097-087 and CGCG~097-092.
As the tail of CGCG~097-087 
points opposite to the direction to the 
cluster center (Figure \ref{fig:FC1}), it is natural to think that 
CGCG~097-087 would be entering the cluster directly to the center 
on the tangential plane.
Meanwhile, the tail of 097-092 shows no alignment
toward the cluster center (Figure \ref{fig:FC1}).

\subsection{Morphology of the Cloud-Parent Connection}
\label{sec:morphtype}

As in \citet{Yagi2010}, the EIGs of \A1367 are classified into three types;
(1) connected \Ha\ clouds with a disk-wide star formation in the parent, 
(2) connected \Ha\ clouds without a disk-wide star formation in the parent,
and (3) detached \Ha\ clouds.
Type-1 includes CGCG~097-073, CGCG~097-079, CGCG~097-087, CGCG~097-087N,
CGCG~097-093, CGCG~097-122, 
and three BIG galaxies (CGCG~097-120, CGCG~097-114, and CGCG~097-125).
Type-2 includes 2MASX~{\allowbreak}J11443212{\allowbreak}+2006238.
Though star formation partly remains in the galaxy,
it resembles GMP3779 and GMP3896 in \citet{Yagi2010}, 
which were classified as type-2.
Type-3 includes CGCG~097-092, SDSS~J114501.81{\allowbreak}+194549.4, and the orphan clouds.
The type classification is given in Tables \ref{tab:clouds} and \ref{tab:parents}.

In the Coma cluster, the number of each type was
4 (Type 1), 4 (Type 2), and 6 (Type 3).
In \A1367, the fraction of Type 1 is larger and Types 2 and 3 are smaller.
In the Coma cluster, four faint blue parents with detached clouds (Type 3)
are found, and they show post-starburst spectrum. 
In \A1367, there are fewer post-starburst parents.
Only SDSS~J114501.81+194549.4 in BIG shows a sign of post-starburst signature,
whose color is barely red (0.01 mag redder than the demarcation line).

\section{Discussion}

\subsection{The difference in EIGs and parents in \A1367 and in Coma}
\label{sec:comparison}

Though many statistical properties of EIGs and their parents 
are similar in \A1367 and the Coma, there are several differences.
In \A1367 the EIGs are longer, and tend to be connected to parents.
The EIG parents in \A1367 have more star formation in the disk,
and post-starburst ones are fewer.
The fraction of EIG parents and blue galaxies is higher in \A1367.

Under the assumption that the EIG-parents connection 
would evolve from Type 1 to Type 3, 
the larger number of Type 1 in \A1367 and fewer Type 3 suggest that 
the EIGs and parents in \A1367 is in an earlier stage
of the evolution on average.
Most of EIG parents are thought to be galaxies which 
infell into the clusters. 
The gas in the galaxy will be lost due to certain processes,
while part of the gas may form an EIG.
After some time, the star formation in the galaxy will cease,
and the color of the galaxy will eventually become redder.
The larger fraction of blue galaxies in \A1367 suggests
that not only the EIG parents but also other galaxies 
are still unprocessed.
As the dominant mechanism of the gas removal 
in cluster environment
is thought to be a ram-pressure stripping
\citep[e.g.,][]{Boselli2016}, 
it is natural that that the evolution is slower in \A1367.
From the X-ray observation by ASCA \citep{Fukazawa2004},
the central gas density of \A1367 is about 1/5 of that of the Coma,
the mass is about 1/14, and the core radius is 30\% larger.
Thus the ram pressure, which is
proportional to the density and the square of the velocity,
should be weaker in \A1367.
The smaller number of parents with post-starburst signature
in \A1367 is also explained by the lower efficiency 
of the ram-pressure stripping.

The longer length of EIGs in \A1367 requires another explanation.
The length of EIGs shows the distribution of the ejected 
gas from the parents and/or the reach of the ionization photons.
If the ionization mechanism is something that is spatially restricted,
such as an UV from the young stars of the parent galaxy or inside of EIG,
or AGN, the EIG length would be determined by the range of the flux.
If it is the case, however, a gradient of SB 
according to the distance from the UV source is expected, 
which is not observed in the EIGs in \A1367 and in the Coma.
Therefore the length would mainly represent the distribution 
of the ionized gas.
According to GOLDMine \citep{Goldmine},
the parent galaxies of a long EIG in \A1367
have a stellar mass comparable to those in the Coma.
Thus the original gas before stripping would have been comparable, too.
We present two possible explanations on the longer distribution 
of the ionized gas in \A1367.
One is that the removal of the gas is relatively 
slower and longer lasting in \A1367,
because of the low efficiency of the ram-pressure stripping.
Another possibility is that the short length of EIGs in the Coma 
does not mean the short duration of the gas removal, but 
the distant part of the gas was heated up and got invisible in \Ha
\citep{Tonnesen2010,Tonnesen2011}.
The length reflects the survival time scale of the ionized gas 
in the cluster.
Though the uncertainty is large, 
the estimated masses of the ionized gas of EIGs in \A1367\ 
are larger than those in the Coma by about an order,
while the difference in the length of EIGs is about twice.
The mean surface brightness in \Ha\ is also brighter 
in EIGs in \A1367.
These results suggest that the slow heating scenario
would be more likely.

\subsection{Star-forming Blobs in or near EIGs}
\label{sec:starformation}

There are several studies on
star formation in stripped tail;
inside BIG \citep{Gavazzi2003}, 
and galaxies in other clusters \citep[e.g.][]{
Sun2007,Sun2010,Cortese2007,
Yoshida2008,Hester2010,Yoshida2012,Yagi2013b,Kenney2014}.
Results from simulations are also reported
\citep[e.g.,][]{Kronberger2008,Kapferer2008,Kapferer2009,
Tonnesen2010,Tonnesen2011,Tonnesen2012}.
Regarding the tail of BIG,
\citet{Cortese2006} predicted that star formation may occur
since the mean column density is high enough to start star formation.

In the B, R, and \NB composite image in this study, 
we detected possible SF blobs in the tail of BIG and near the orphan clouds
(Figures \ref{fig:orphans},\ref{fig:BIGSFR},
and Table \ref{tab:SFRs}.)
We checked the GALEX image from the archive, and found that 
they are also detected in UV.
We refer to the candidates as SF blobs for simplicity, hereafter,
though spectroscopic confirmation is needed to confirm
that they are genuine SF blobs in \A1367.
We named SF blobs in the tail of BIG B1 -- B6, 
according to the distance from BIG,
and the one near the orphan clouds (orphan2) as O1.
Their properties are shown in Table \ref{tab:SFRs},
and cutouts are shown as Figures \ref{fig:BIGSFR_stamp}
and \ref{fig:orphansSFR_stamp}.
In this section, we mainly used images before PSF matching 
to make use of higher spatial resolution in NB band.
Meanwhile the magnitude shown in Table \ref{tab:SFRs}
was measured in PSF matched images for consistency.
The magnitude and \Ha\ flux were measured within the isophote of
2.5$\times 10^{-18}$ erg s$^{-1}$ cm$^{-2}$ arcsec$^{-2}$.
We assumed z=0.0275 for B1--B6, and z=0.0217 for O1,
respectively.
In B, R, and \NB composite of B2, B4, and B5,
possible blended fore/background objects are 
recognized as yellow color objects.
Thus their magnitudes are contaminated by the blend.
Also it should noted that since the PSF and deblending is different,
the magnitude in optical and UV is not directly comparable.

The regions are very blue in UV (FUV-NUV) and in optical (B-R and R-i);
$-1.4<$FUV-NUV$<0.6$,
$-0.3\leq$B-R$\leq0.6$, and
$-0.5\leq$R-i$\leq0.3$, before internal extinction correction.
The color implies that they are SF blobs and 
do not suffer heavy dust extinction.
The \Ha\ luminosity of SF blobs is estimated to be
2--7$\times 10^{36}$ erg s$^{-1}$,
though the value has quite large uncertainty
as discussed in Section \ref{sec:error}.

\subsubsection{SF blobs in the tail of BIG }

The tangential distance of the SF blobs from galaxies in BIG is
63--150 kpc from CGCG~097-120, 120--190 kpc from CGCG~097-114,
or 110--200 kpc from CGCG~097-125.
The most distant SF blob, B6, is $>$ 150 kpc away.
Even if the tangential motion of the blob relative to BIG is 
as high as 1000 km s$^{-1}$, 
B6 requires $\gtrsim$ 1.5$\times 10^8$ years to reach the distance.
Thus it would not be a stripped star-forming region from the parent,
but in-situ star formation.

In B4, possible offset between B and \Ha\ is recognized;
B-band emission is stronger near BIG.
In Figure \ref{fig:BIGSFR_stamp} top, the blue component is
extended to bottom-left.
They resemble ``fireball'' features found in EIG of 
RB199 in the Coma cluster \citep{Yoshida2008};
the ram pressure works at the regions, and \Ha\ 
emitting clouds were swept downstream while formed young stars were not.
The offset roughly aligns to the tail and 
points in the direction of CGCG~097-120 and CGCG~097-125.
This supports the assumption that B4 is a part of the tail.

\subsubsection{SF blobs near orphan2}

The SF blobs near orphan2 align almost perpendicularly to the
\Ha\ filament from orphan2 (Figure \ref{fig:orphansSFR}).
Though SF blobs are resolved into several compact sources 
in B and \NB images before smoothing, 
they are merged in PSF matched images,
and we measured the magnitude as a whole in Table \ref{tab:SFRs}.

In Figure \ref{fig:orphansSFR_stamp} right, 
several residuals remain in \Ha\ image.
Figure \ref{fig:orphansSFR_stamp} center, B, R, and i composite is 
shown as a reference; the objects that show orange--yellow color 
in B, R, and i composite are not SF blobs, and the \Ha\ excess 
are fake, because of PSF mismatch, and disappear in PSF matched \Ha\ image.

In broadband images, a faint stream is seen near the SF blobs.
B-band image is shown as \ref{fig:orphansSFR} bottom-left.
Because \Ha\ is weak along the stream (Figure \ref{fig:orphansSFR_stamp})
the stream would be stars and might be a tidal tail.
The stream seems to connect two galaxies,
SDSS~J114425.39{\allowbreak}+200923.2,
and SDSS~J114426.00{\allowbreak}+201001.8
across the orphan2 filament.
Their projected distance is 18.3 kpc, if they are at the \A1367 distance.
At the opposite side of 
SDSS~J114426.00{\allowbreak}+201001.8,
the stream is vague but possibly reaches to the UDG,
which is about 34 kpc away from the south galaxy (SDSS~J114425.39{\allowbreak}+200923.2).
As the three galaxies,
SDSS~J114425.39{\allowbreak}+200923.2,
SDSS~J114426.00{\allowbreak}+201001.8,
and the UDG does not have spectroscopic information,
the galaxies may not be an \A1367 member.
Meanwhile, as the SF blobs show strong excess in \Ha, 
they would be in \A1367.

As the SF blobs align to the blue stream perpendicular to 
the \Ha\ filament of orphan2, and they are 
4--8 kpc away from the \Ha\ filament,
they would be an accidental overlap on the orphan clouds.
In Table \ref{tab:clouds}, the flux of orphan2 therefore 
does not include the SF blobs.

\section{Summary}

We investigated \Ha\ images of \A1367 taken with Suprime-Cam,
and made a catalog of extended ionized gas clouds (EIGs) and their parents.
Though \A1367 is one of the best-studied clusters, we added six new EIGs
to enable a statistical discussion and a comparison with EIGs in the 
Coma cluster.
The deep \Ha\ image also revealed that 
the \Ha\ tails are extended in fainter surface brightness
much longer than previously known.
The tail of blue infalling group(BIG) shows a sign of winding
which could be a result of motion of galaxies inside the group.
We also found several candidates of star-forming blobs
(SF blobs) far from parent galaxies.
The comparison of the parent galaxies of EIGs in 
\A1367 and in the Coma cluster showed that 
the properties of the parents are basically similar.
Meanwhile the length of EIGs is longer and 
more often connected to star-forming parents in \A1367.
The difference suggests that the EIGs and parents in \A1367 are,
on average, younger than those in the Coma cluster,
and in \A1367, the gas removal from the parent 
and/or the heating of EIG is slower.
It would reflect a different evolutionary stage 
of the clusters.

\acknowledgments
We thank the anonymous referee for thorough reading and 
important comments and suggestions.
We thank Subaru Telescope staff for their help during the observation.
This work made extensive use of the GOLDmine database
\footnote{\url{http://goldmine.mib.infn.it/}}
\citep{Goldmine,Gavazzi2014}.
This work has made use of the 
SDSS database \footnote{\url{http://cas.sdss.org/}}, 
NED database,
Mikulski Archive for Space Telescopes (MAST),
and computer systems at Astronomical Data Analysis Center of 
National Astronomical Observatory of Japan(NAOJ).
We acknowledge Dr. Michele Fumagalli for the information
about SDSS~J114513.76+194522.1 from his recent observation,
and Guido Consolandi for the quick reduction of the data.
This research is partly supported by
MEXT KAKENHI 24103003 and JSPS KAKENHI 15H02069.

\onecolumn

\begin{figure}[ht]
\includegraphics[scale=5,bb=0 0 45 64]{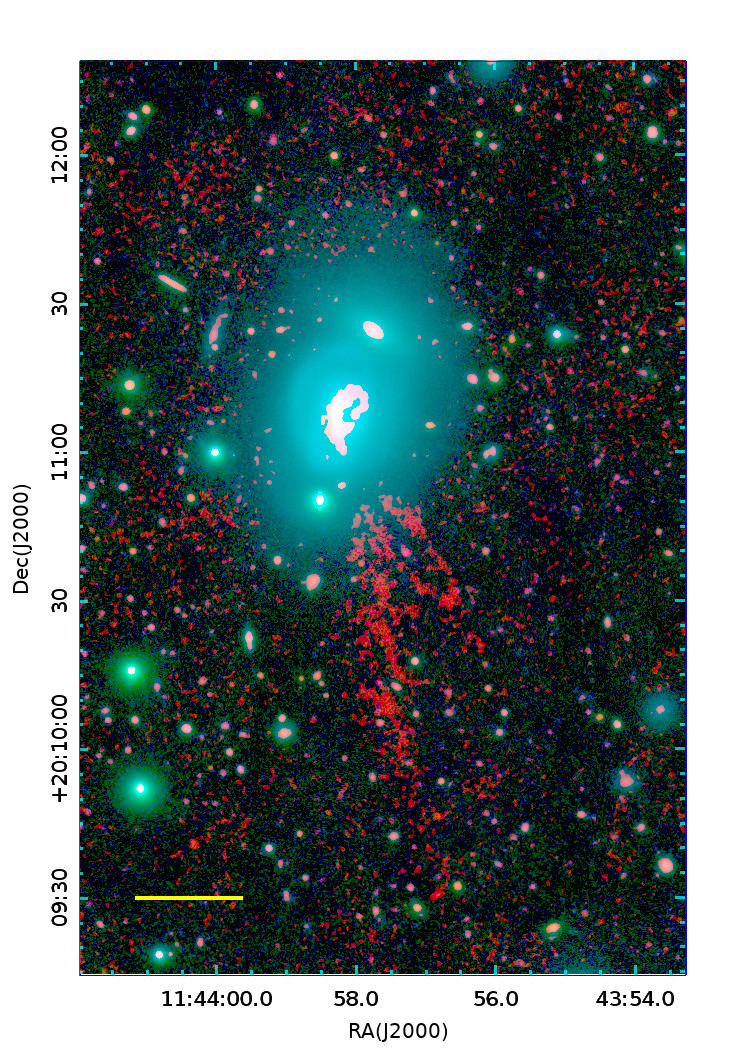}
\includegraphics[scale=5,bb=0 0 45 64]{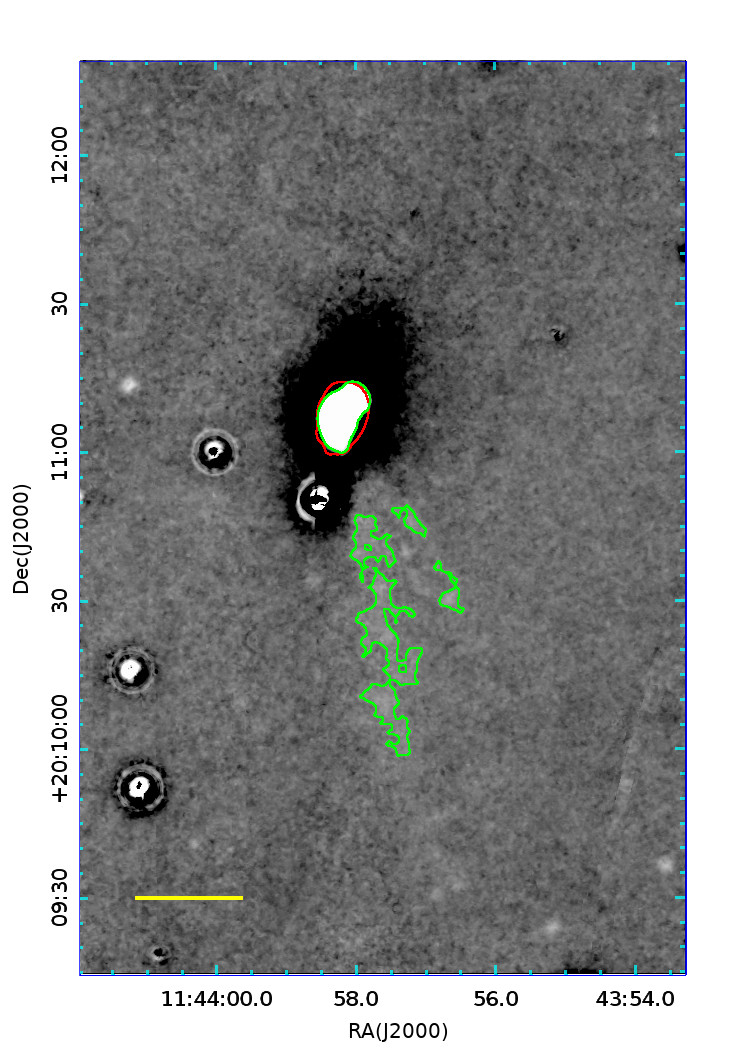}\\
\caption{Same as Figure \ref{fig:orphans} top panels, 
but around CGCG~097-092. In the right panel,
the redshift in Table \ref{tab:parents} is used for 
converting the \Ha\ surface brightness to the isophote.
The red contour represents the isophote of the R-band image
which corresponds to the SDSS r-band $petroR50$ (the radius 
containing 50\% of the Petrosian flux).
}
\label{fig:097092}
\end{figure}

\begin{figure}[ht]
\includegraphics[scale=8,bb=0 0 26 38]{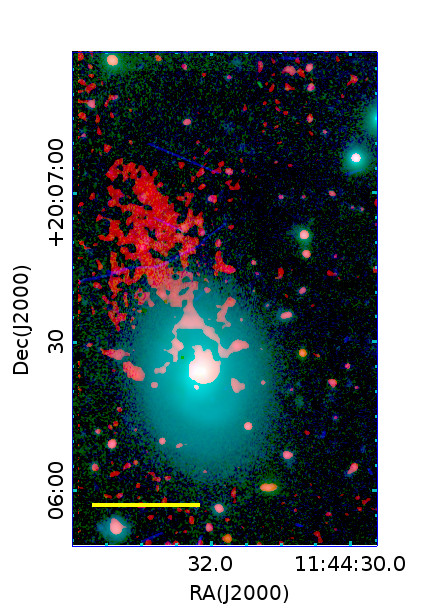}
\includegraphics[scale=8,bb=0 0 26 38]{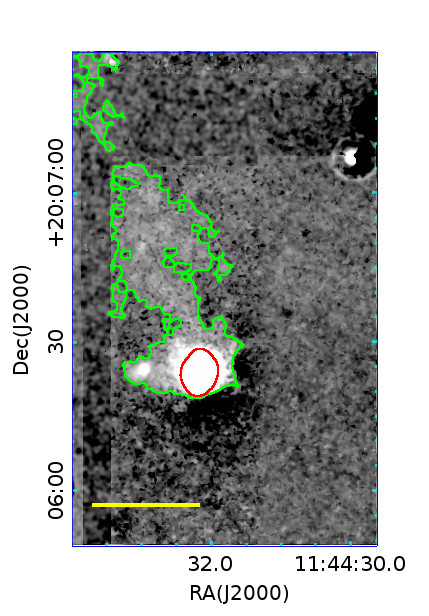}\\
\caption{Same as Figure \ref{fig:097092},
but around 2MASX~J11443212+2006238.
}
\label{fig:2MASXJ}
\end{figure}

\begin{figure}[ht]
\includegraphics[scale=4.8,bb=0 0 51 52]{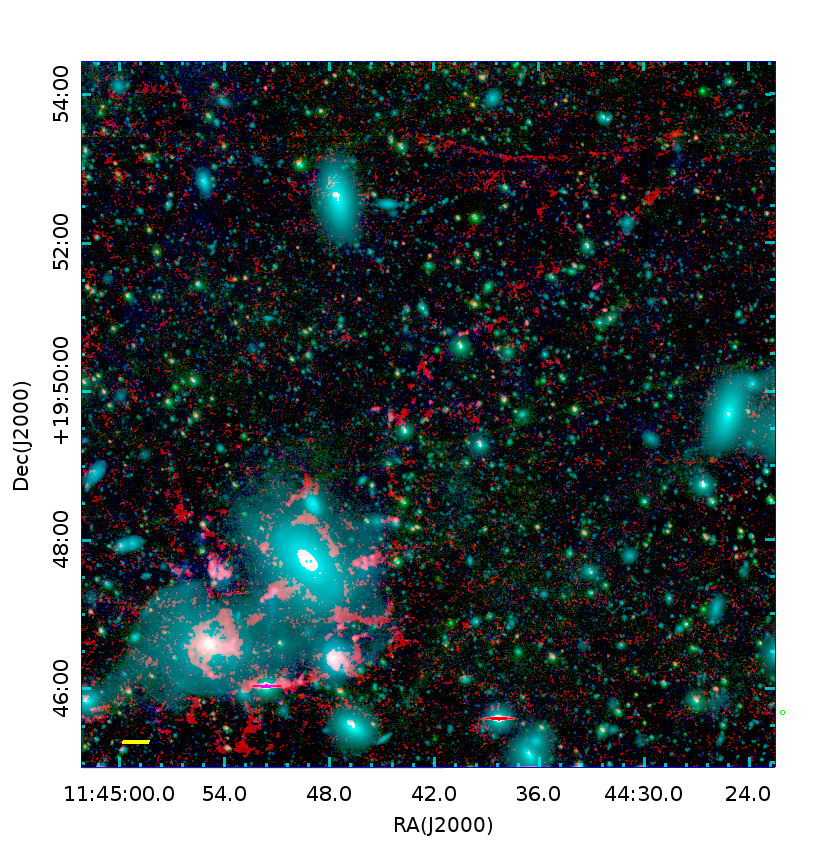}
\includegraphics[scale=4.8,bb=0 0 51 52]{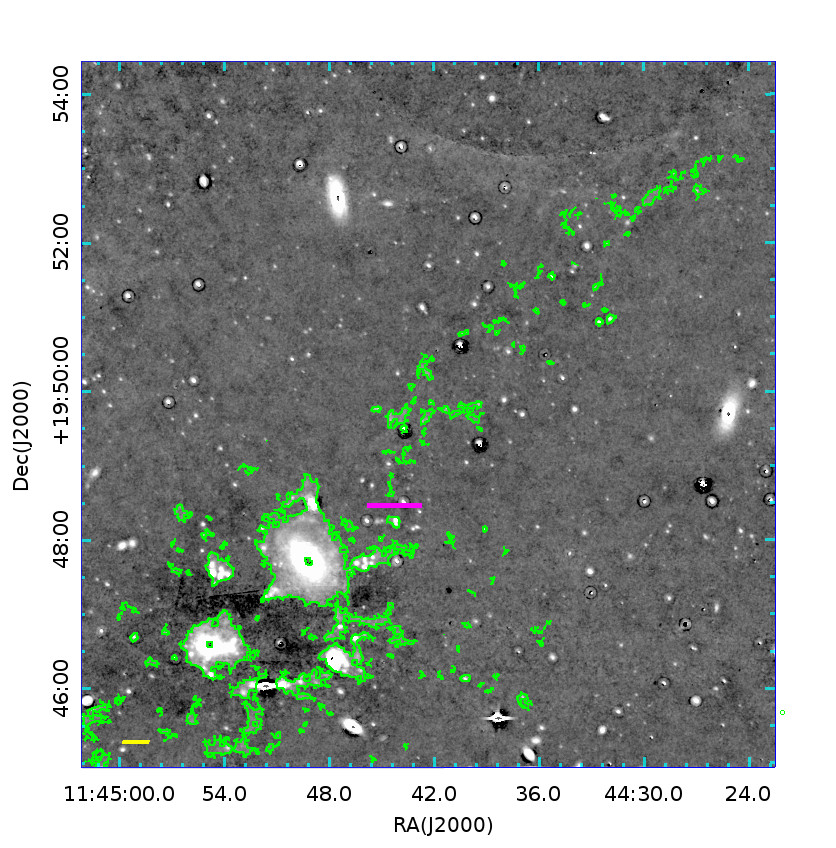}\\
\includegraphics[scale=4.8,bb=0 0 51 52]{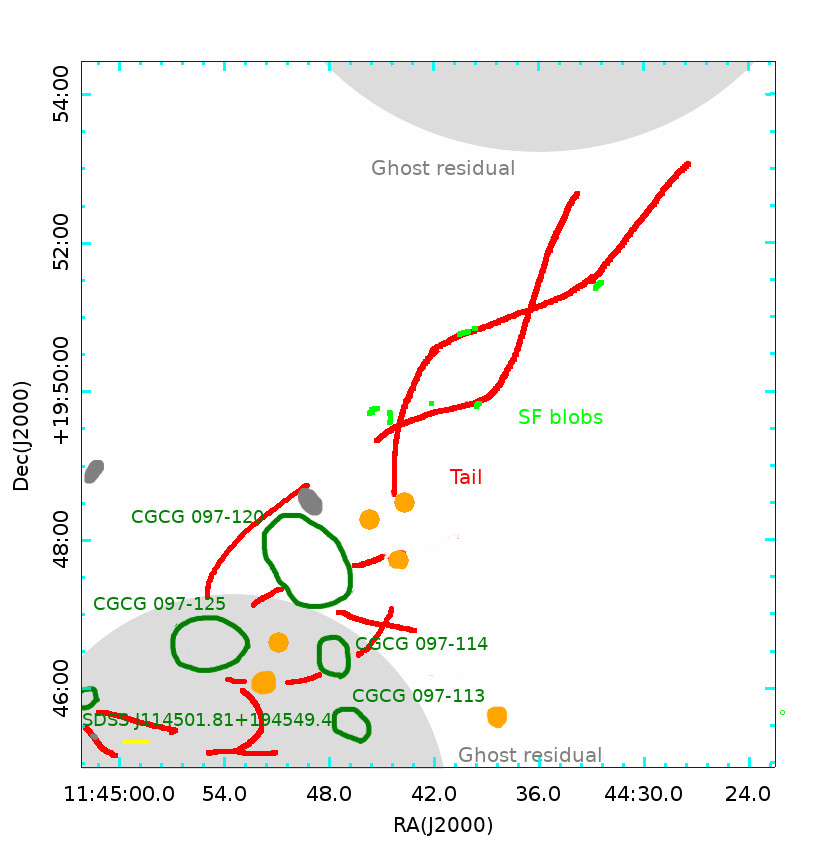}
\includegraphics[scale=4.8,bb=0 0 51 52]{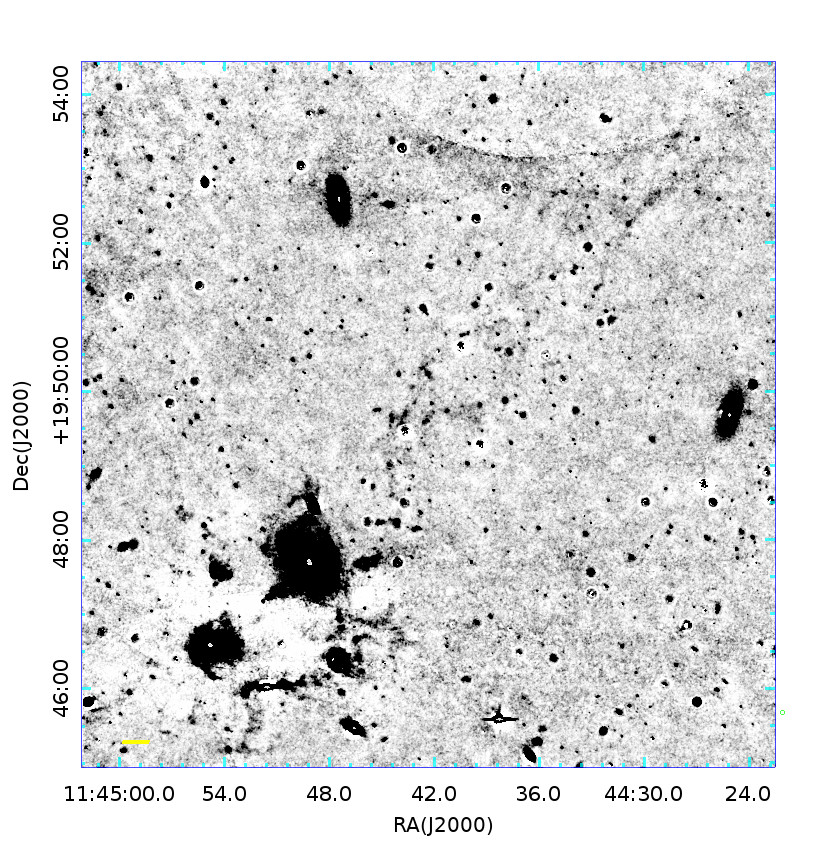}
\caption{(Top panels)
Same as Figure \ref{fig:097092}, but around BIG and its tail.
In the top-right panel,
z = 0.0275($v_r$ = 8244 km s$^{-1}$) is assumed for the green 
contour.
Magenta line shows the adopted boundary of BIG and the tail
for flux measurement.
(Bottom-left)
Schematic figure of BIG and tail is shown.
Green ellipses are member galaxies of \A1367
and gray ellipses are background galaxies.
Orange-filled circles are stars.
Only objects near BIGs are drawn.
Heavy ghost residuals are also shown as light gray circles.
The possible winding streams are depicted in red.
Possible SF blobs are shown in light green.
(Bottom-Right) Inverted color image of \Ha, with different 
contrast to show the morphology of the tail of BIG clearer.
}
\label{fig:BIGtail}
\end{figure}

\begin{figure}[ht]
\includegraphics[scale=3.5,bb=0 0 89 56]{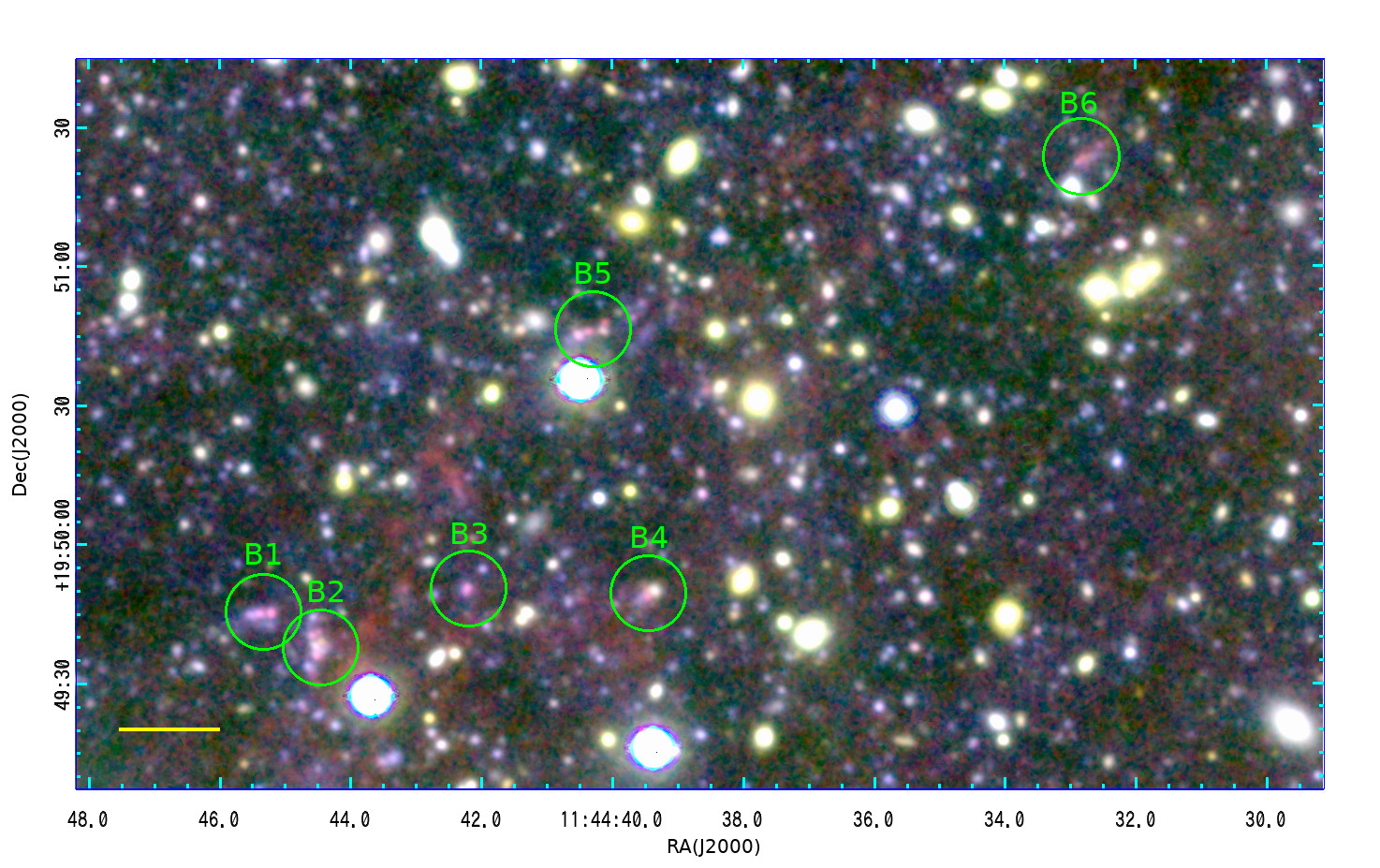}\\
\includegraphics[scale=3.5,bb=0 0 89 56]{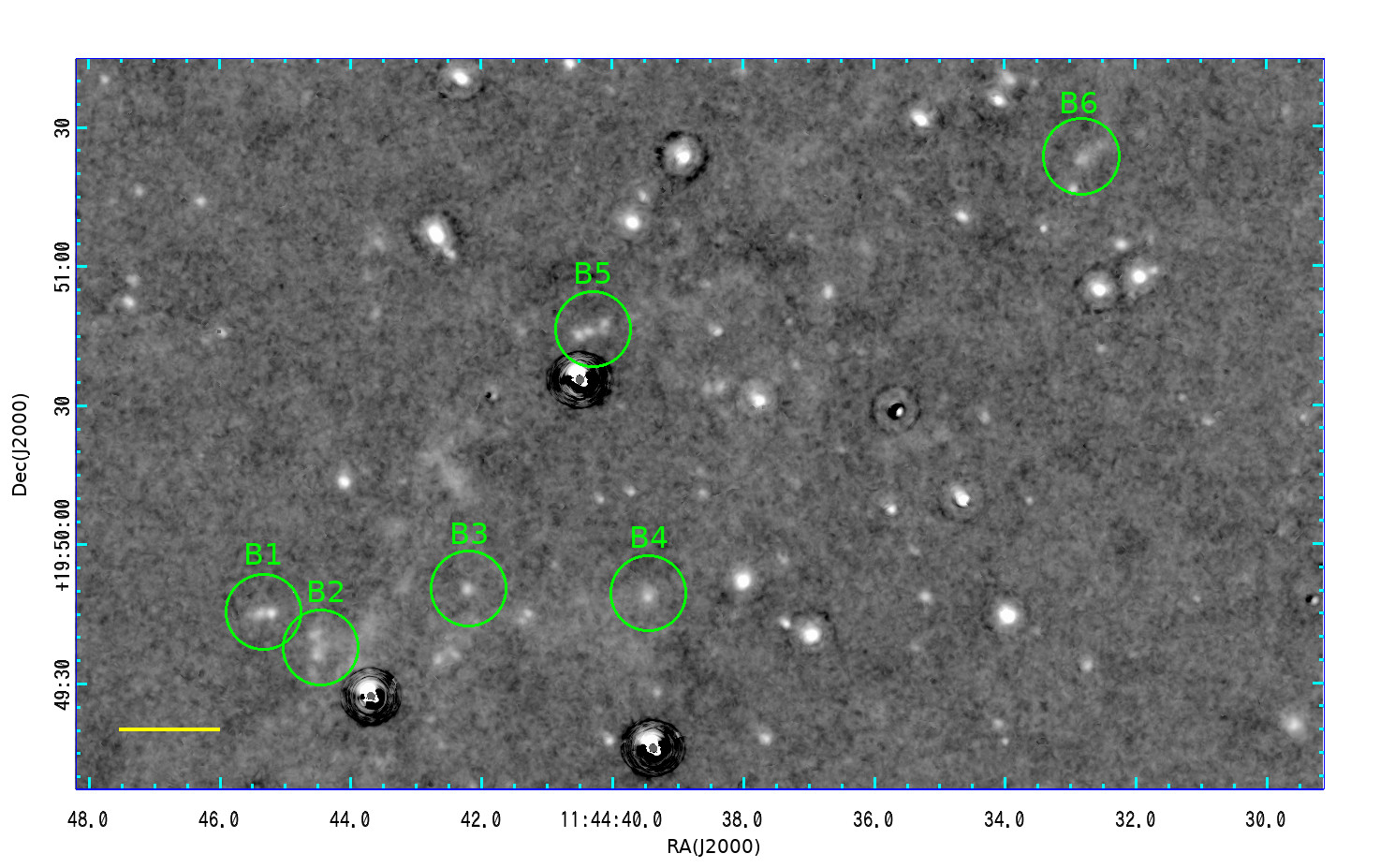}\\
\includegraphics[scale=3.5,bb=0 0 89 56]{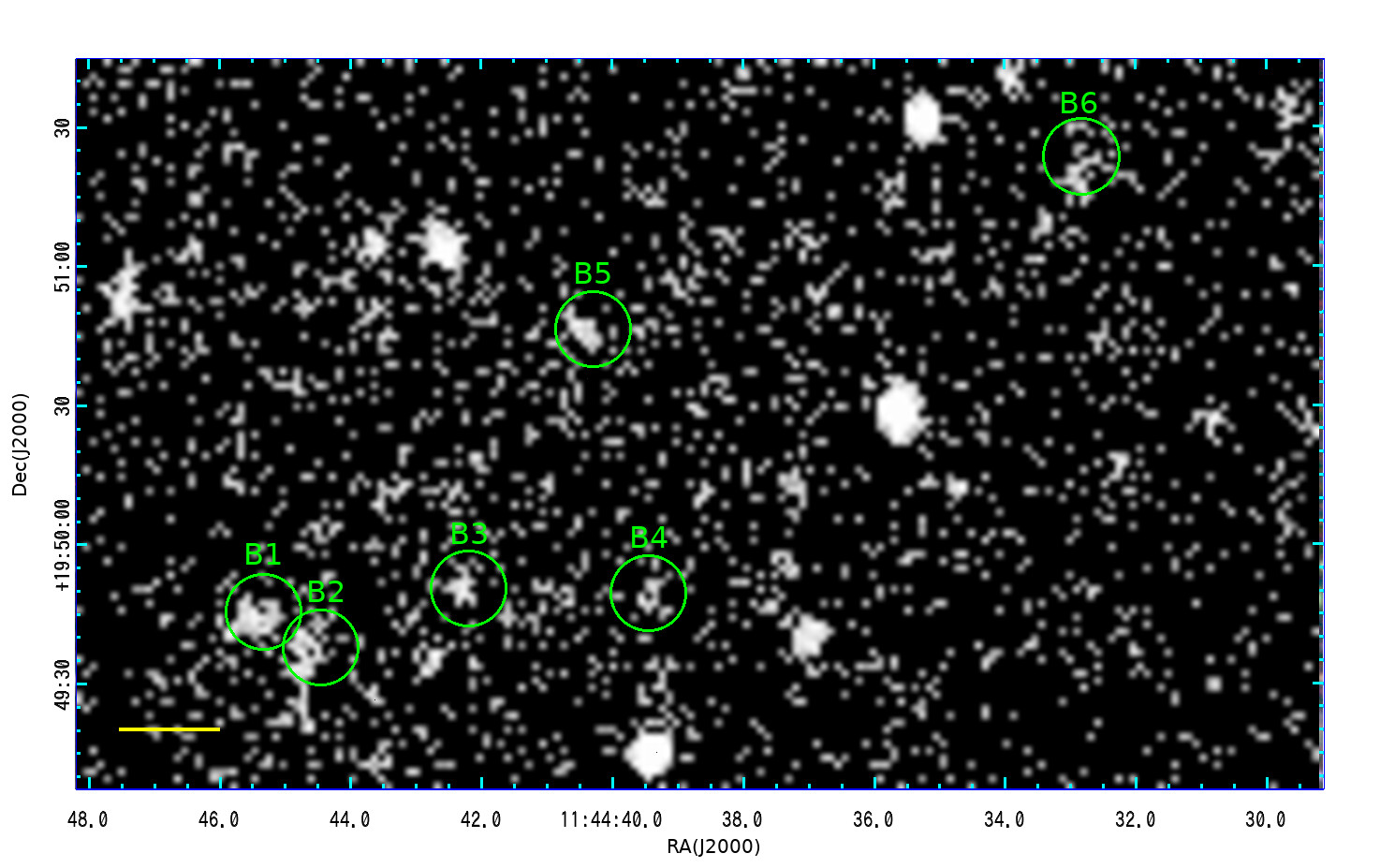}
\caption{
Zoomed-up images around the SF blob candidates in the tail of BIG.
Six SF blob candidates are designated as green circles.
From the top to the bottom,
B, R, and \NB composite, \Ha, and NUV image from GALEX archive 
are shown.
}
\label{fig:BIGSFR}
\end{figure}

\begin{figure}[ht]
\includegraphics[scale=4,bb=0 0 56 61]{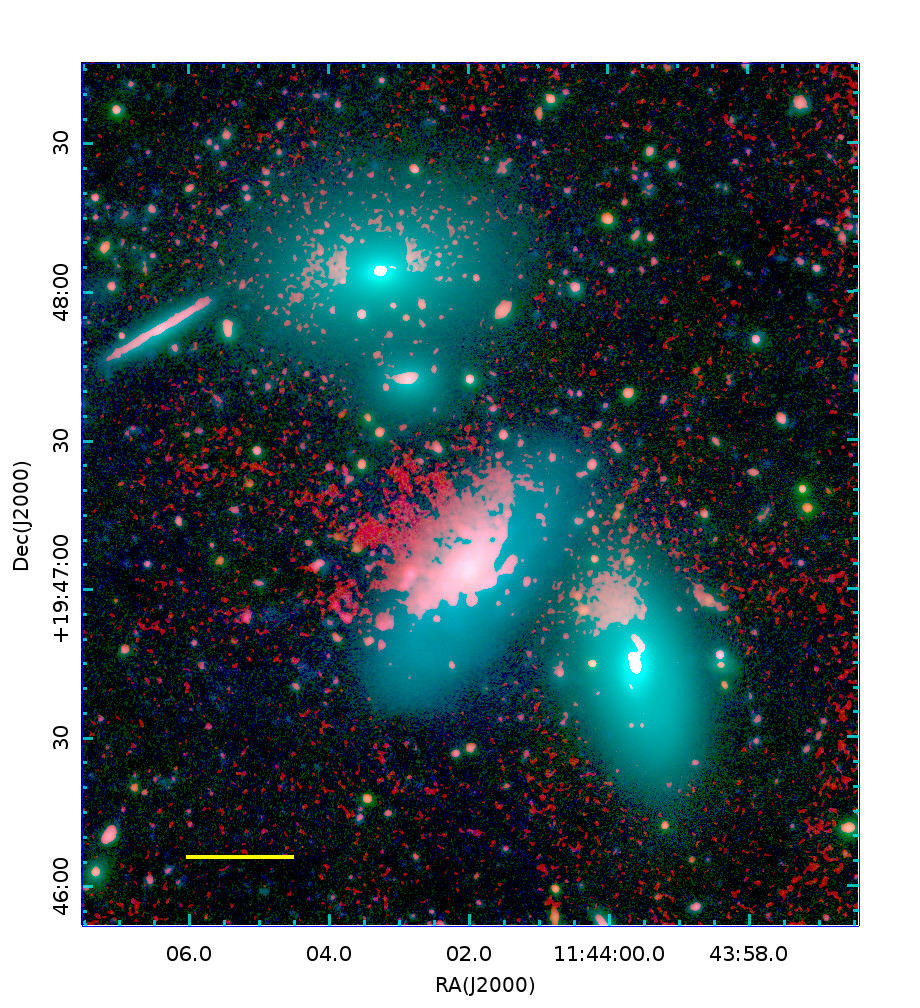}
\includegraphics[scale=4,bb=0 0 56 61]{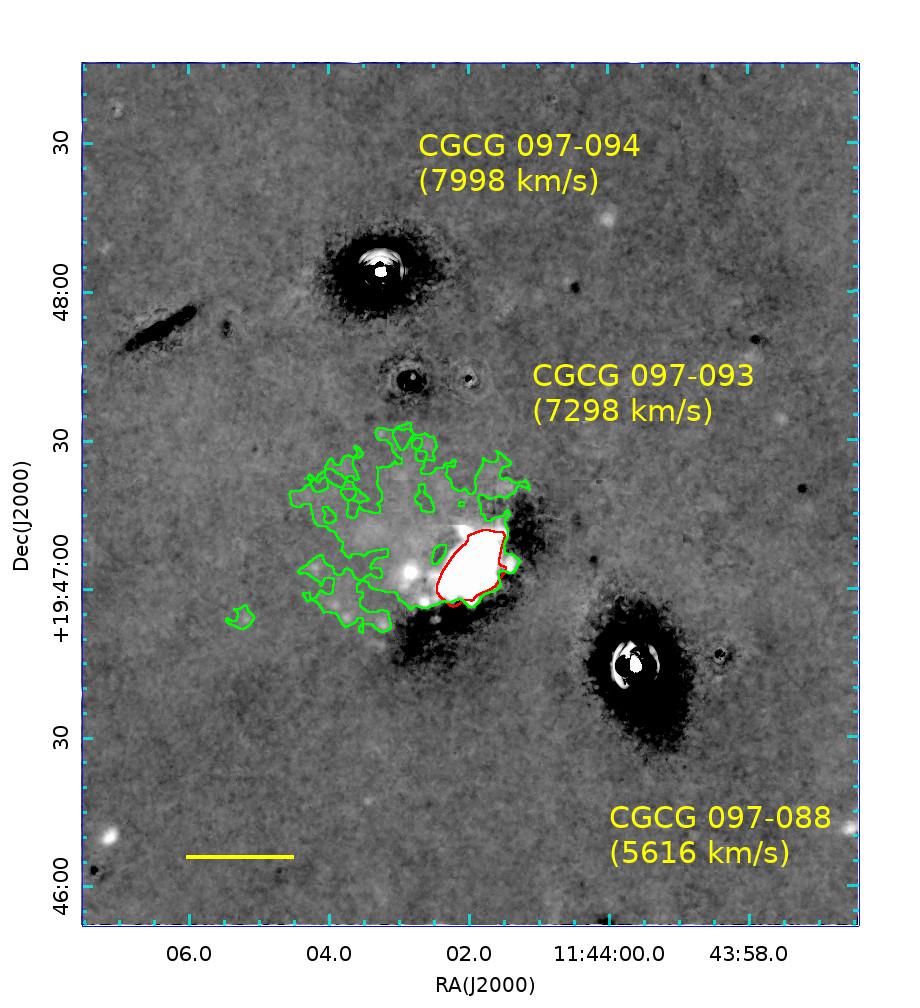}\\
\caption{
Same as Figure \ref{fig:097092}, but around CGCG~097-093.
Recession velocities of three galaxies are also shown.
}
\label{fig:097093}
\end{figure}

\begin{figure}[ht]
\includegraphics[scale=3.5,bb=0 0 68 61]{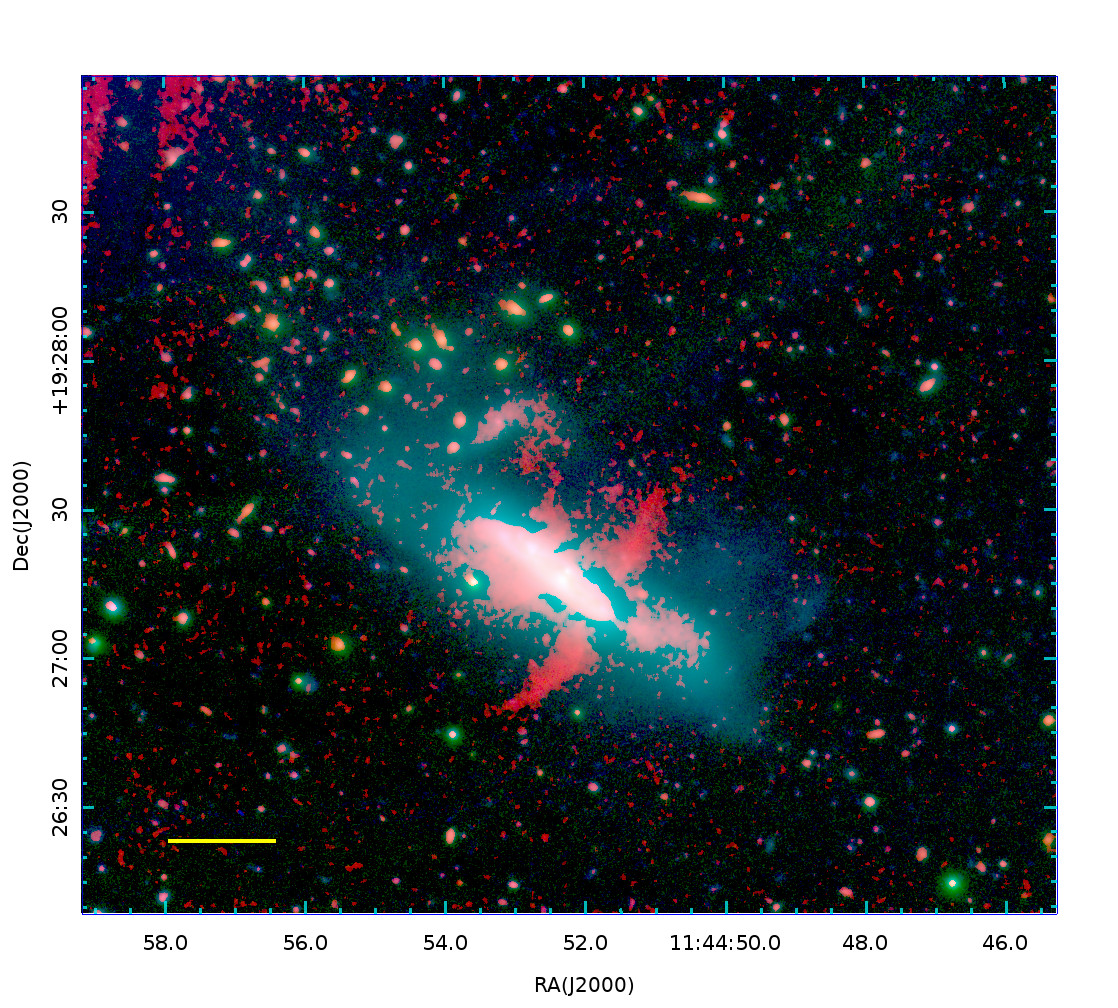}
\includegraphics[scale=3.5,bb=0 0 68 61]{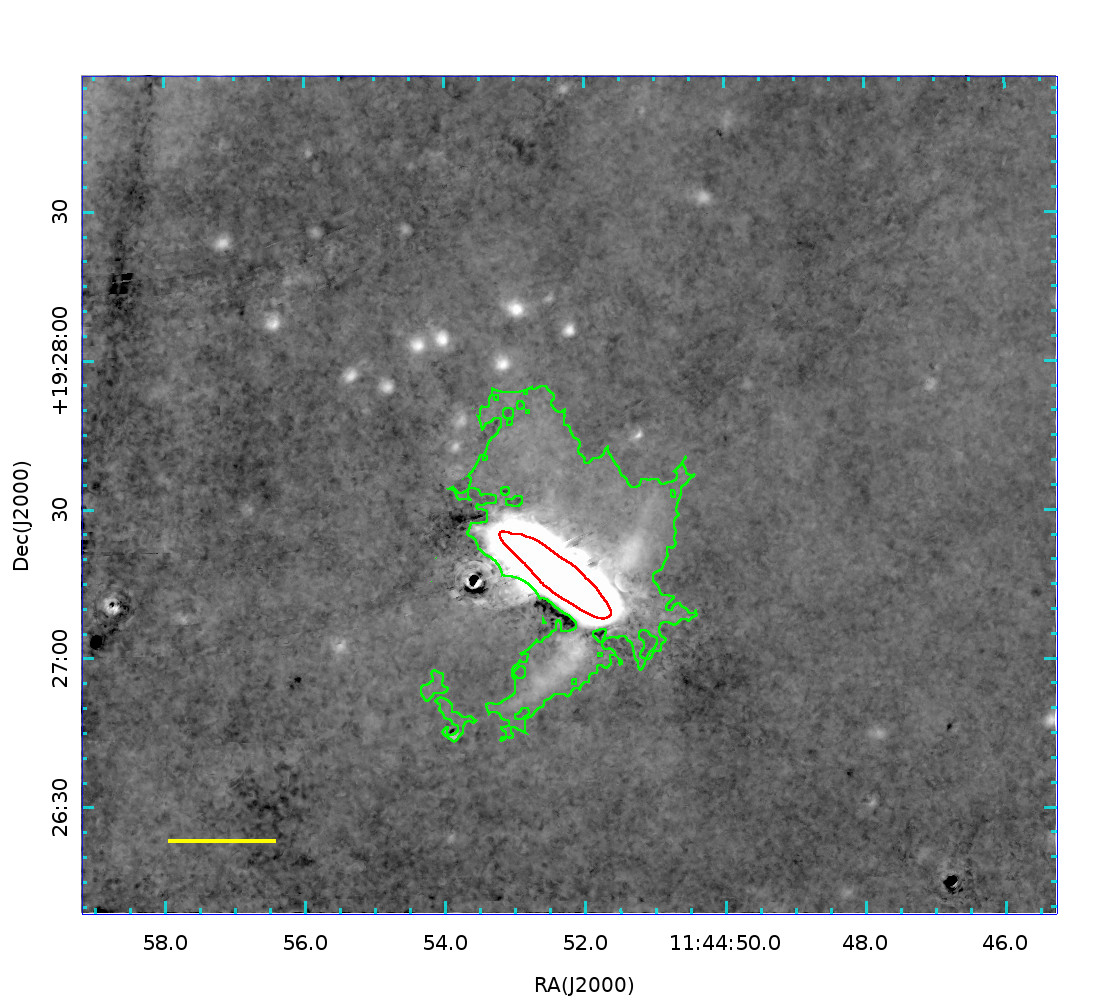}\\
\caption{Same as Figure \ref{fig:097092}, but around CGCG~097-122.
}
\label{fig:097122}
\end{figure}

\clearpage

\begin{figure}[htb]
\includegraphics[scale=0.8,bb=0 0 574 574]{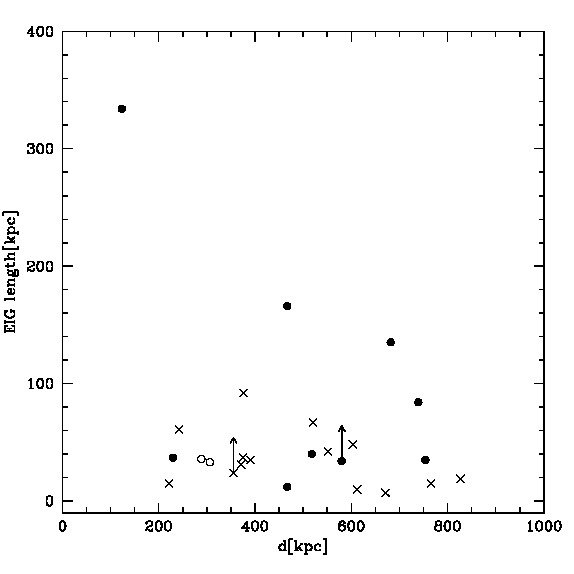}
\vspace{3mm}
\caption{
Cluster-centric projected distance versus the length of EIG.
Filled circles are EIGs of \A1367, and xs are those of the Coma.
Open circles are the orphan clouds.
The cluster-centric distance is measured at the position of 
the parent galaxy except the orphan clouds.
The vectors show GMP3816 and 2MASX~J11443212+2006238,
as their EIGs extend out of the observed area.
}
\label{fig:length}
\end{figure}

\clearpage

\begin{figure}[htb]
\includegraphics[scale=0.8,bb=0 0 574 574]{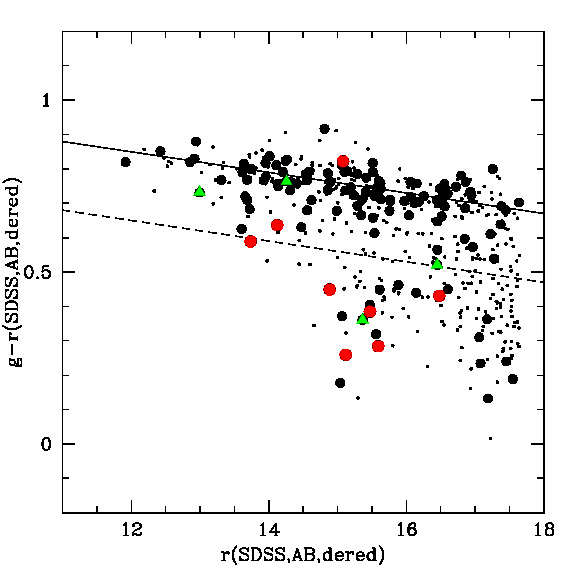}
\vspace{3mm}
\caption{Color magnitude diagram of member galaxies of \A1367.
The color and magnitude data are taken from SDSS DR12
except for CGCG~097-087, whose color and magnitudes are 
supplemented by GOLDMine Database \citep{Goldmine},
as described in the Appendix.
Galaxies in the observed field and those out of the field
are shown as filled circles and points, respectively.
The red circles show parents of extended \Ha\ clouds.
Four BIG members 
(CGCG~097-120, 114, 125, and SDSS~J114501.81~+194549.4)
are marked by green triangles.
The solid line shows color magnitude relation (CMR) 
of early type galaxies, and the broken line is 0.2 mag
bluer than the CMR for demarcation of blue and red galaxies.
}
\label{fig:CMD}
\end{figure}

\begin{figure}[ht]
\includegraphics[scale=0.8,bb=0 0 574 574]{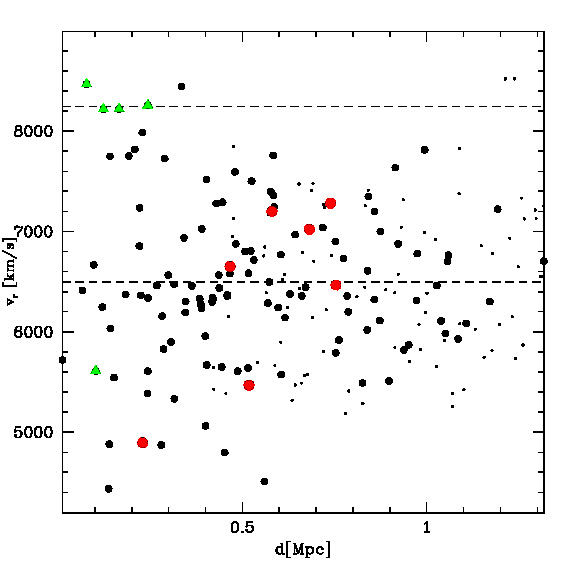}
\caption{Distance from the cluster center versus recession velocity plot.
The data are taken from SDSS DR12.
The symbols are the same as Figure \ref{fig:CMD}.
Horizontal broken lines show the recession velocity of \A1367 
(v = 6494 km s$^{-1}$) and BIG (v = 8244.3 km s$^{-1}$).
}
\label{fig:vd}
\end{figure}

\begin{figure}[ht]
\includegraphics[scale=0.67,bb=0 0 666 222]{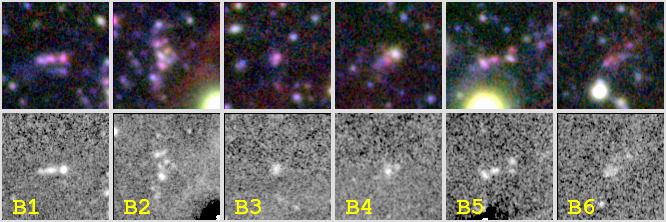}
\caption{
SF blob candidates in the tail of BIG.
For high spatial resolution images before PSF matching are used.
From the left to the right,
B, R, and NB composite(top) and \NB-R(bottom) 
of B1, B2, ... and B6 are shown.
The size is 10 kpc square.
}
\label{fig:BIGSFR_stamp}
\end{figure}

\begin{figure}[ht]
\includegraphics[scale=1.0,bb=0 0 333 111]{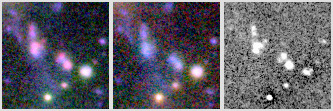}\\
\caption{
Cutout of SF blob candidate near orphan2.
The size is 10 kpc square, and images before PSF matching are used. 
From the left to the right,
B, R, and NB composite, B, R, and i composite, and \NB-R(\Ha) are shown.
}
\label{fig:orphansSFR_stamp}
\end{figure}

\begin{figure}[ht]
\includegraphics[scale=3.5,bb=0 0 62 71]{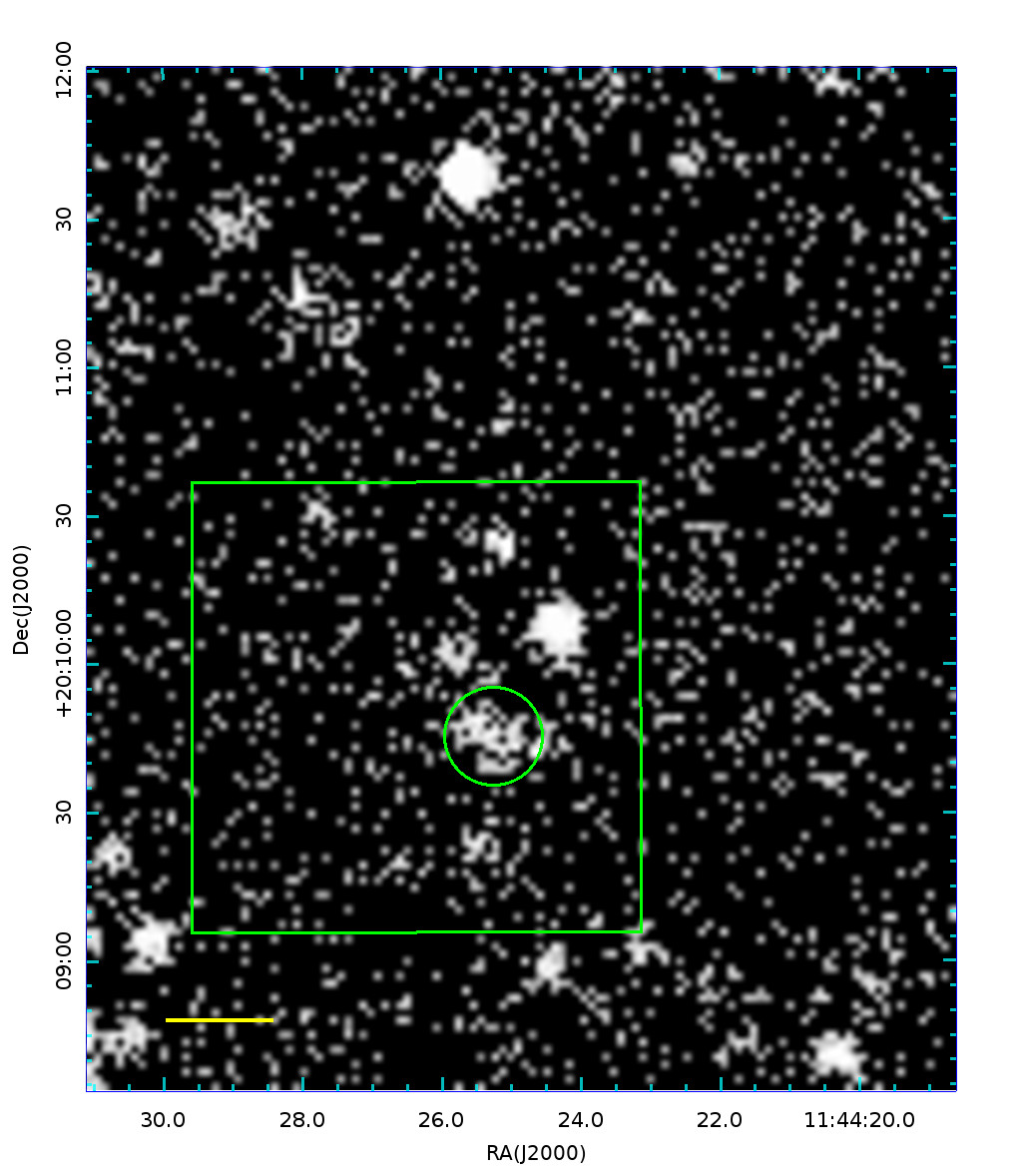}\\
\includegraphics[scale=4.5,bb=0 0 35 34]{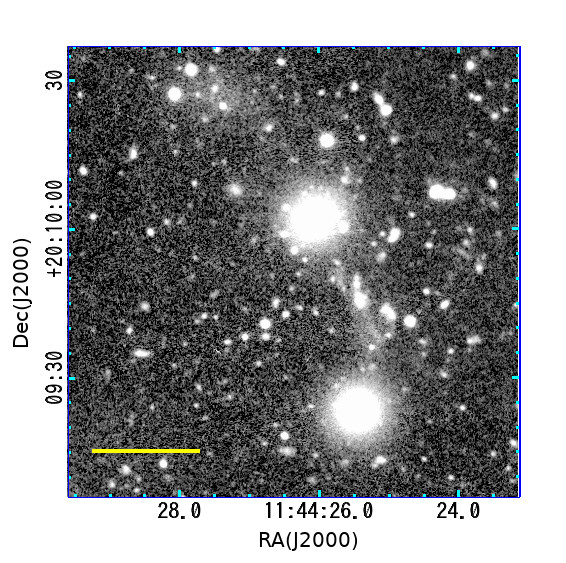}
\includegraphics[scale=4.5,bb=0 0 35 34]{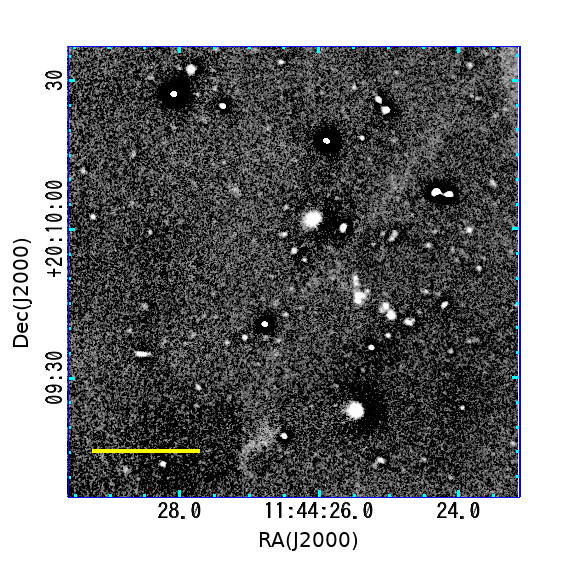}
\includegraphics[scale=4.5,bb=0 0 35 34]{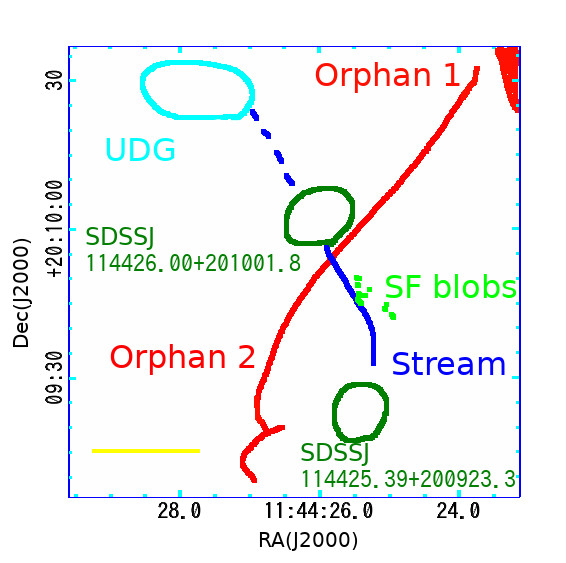}\\
\caption{
(Top) 
GALEX NUV image at the same region as Figure \ref{fig:orphans}.
The green circle indicates the SFR.
The regions of the bottom figures are shown as green squares.
(Bottom Left) B-band image before PSF matching.
(Bottom Center) \Ha(\NB-R)image before PSF matching.
The SF blob is resolved.
The large R-band PSF makes a negative envelope around stars.
(Bottom Right) Schematic figure around the SF blobs.
}
\label{fig:orphansSFR}
\end{figure}

\clearpage
\begin{table}[ht]
\caption{Summary of mosaicked images and observation log}
\label{tab:obs}
\begin{tabular}{|c|c|c|c|c|}
\hline
filter &
PSF size (FWHM) & SB$_{\rm lim}$(ABmag arcsec$^{-2}$)\tablenotemark{a}&
Date(UT) & exposure \\
\hline
B & 0''.7 & 28.5 & 2014-05-01 &  8$\times$600sec  \\
\hline
R & 1''.2   & 27.9 & 2014-04-30   & 5$\times$60 sec\\
    &        &      & 2014-04-30 &  10$\times$300  sec\\
    &        &      & 2014-05-01 &   5$\times$300  sec\\
\hline
i & 1''.1 & 27.4 & 2014-05-01 & 10$\times$180 sec  \\
\hline
\NB & 0''.7--0''.9 & 27.7 & 2014-04-30 &  15$\times$900 sec \\
    &        &      & 2014-05-01 &   5$\times$900  sec \\
\hline
\end{tabular}
\tablenotetext{a}{SB$_{\rm lim}$ represents 
1-$\sigma$ fluctuation of 
the surface brightness measured in random apertures
of 2 arcsec diameter.}
\end{table}

\begin{table}[ht]
\caption{Extended \Ha\ Clouds in \A1367}
\label{tab:clouds}
\begin{tabular}{|l|c|c|c|c|c|c|c|}
\hline
ID/parent     & size & area & $\langle$SB$\rangle$ & redshift
 & \Ha\ flux & mass & type \\
(1)&(2)&(3)&(4)&(5)&(6)&(7)&(8)\\
\hline
orphan1       & 33$\times$13 & 128 & 26.6$\pm$0.1& 0.0217 \tablenotemark{c} & (3.4$\pm$0.2)$\times 10^{-15}$ \tablenotemark{c} & (5.4$\pm$0.3)$\times 10^{8}$ \tablenotemark{c}& 3\\
orphan2\tablenotemark{d}       & 36$\times$5 & 27   & 26.7$\pm$0.1& 0.0217 \tablenotemark{c} & (4.0$\pm$0.2)$\times 10^{-16}$ \tablenotemark{c} & (3.1$\pm$0.2)$\times 10^{7}$ \tablenotemark{c}& 3\\ 
\multicolumn{1}{|l|}{2MASX~J11443212} &&&&&&&\\
~+2006238 & $>$33$\times$13\tablenotemark{b}  & $>$177\tablenotemark{b}  
                                   & 23.9$\pm$0.3 & 0.02402 & $>$(6$\pm$2)$\times 10^{-14}$ \tablenotemark{b} & $>$(2.8$\pm$0.9)$\times 10^{9}$ \tablenotemark{b}& 2\\
CGCG~097-092  & 35$\times$11 & 91  & 23.2$\pm$0.3 & 0.02157 & (5$\pm$2)$\times 10^{-14}$ & (1.6$\pm$0.5)$\times 10^{9}$& 3\\ 
~~~~~(without core)& 23$\times$10 & 68  & 26.9$\pm$0.1 & 0.02157 & (1.4$\pm$0.2)$\times 10^{-15}$ & (2.2$\pm$0.3)$\times 10^{8}$& \\ 
CGCG~097-093  & 31$\times$19 & 232 & 24.9$\pm$0.5 & 0.01633 & (4$\pm$3)$\times 10^{-14}$ & (3$\pm$2)$\times 10^{9}$& 1\\ 
CGCG~097-122  & 34$\times$28 & 380 & 23.8$\pm$0.5 & 0.01824 & (1.5$\pm$0.8)$\times 10^{-13}$ & (1.0$\pm$0.5)$\times 10^{10}$& 1\\
\hline         
(BIG+tail)    & 334$\times$145 & 3306 & 25.0$\pm$0.5 & 0.0275\tablenotemark{d} & (5$\pm$3)$\times 10^{-13}$ & (8$\pm$5)$\times 10^{10}$ & \nodata\\
(tail only)   & 184$\times$45 & 486   & 26.7$\pm$0.5 & 0.0275\tablenotemark{d} & (2$\pm$1)$\times 10^{-14}$ & (3$\pm$2)$\times 10^{9}$ & \nodata\\
\hline
CGCG~097-073  & 84$\times$28  & 559 & 24.3$\pm$0.2 & 0.02429 & (1.3$\pm$0.3)$\times 10^{-13}$ & (1.0$\pm$0.2)$\times 10^{10}$& 1\\
CGCG~097-079  & 135$\times$17 & 285 & 23.2$\pm$0.1 & 0.02342 & (1.7$\pm$0.2)$\times 10^{-13}$ & (5.2$\pm$0.6)$\times 10^{10}$& 1\\   
CGCG~097-087  & 166$\times$20 & 910 & 23.1$\pm$0.2 & 0.02218 & (5.8$\pm$1.1)$\times 10^{-13}$ & (2.3$\pm$0.4)$\times 10^{10}$& 1\\ 
CGCG~097-087N & 12$\times$11  & 72  & 24.7$\pm$0.6 & 0.02516\tablenotemark{a} & (1.2$\pm$0.8)$\times 10^{-14}$ & (8$\pm$6)$\times 10^{8}$& 1\\
\hline         
\end{tabular}

(1) Name of the cloud.\\
(2) Size of minimum bounding rectangle of isophote of 2.5$\times$ 10$^{-18}$ erg s$^{-1}$ cm$^{-2}$ arcsec$^{-2}$. Unit is kpc$\times$kpc.\\
(3) Projected area in the isophote. Unit is kpc$^2$.\\
(4) Mean surface brightness in the isophote. Unit is mag arcsec$^{-2}$.\\
(5) Redshift used for calculating \Ha\ flux. Taken from SDSS DR12\citep{DR12} unless otherwise noted.\\
(6) \Ha\ flux. Unit is erg s$^{-1}$ cm$^{-2}$.\\
(7) Mass of the cloud. Unit is $\sqrt{f_v} M_\sun$,
where $f_v$ is a volume filling factor.\\
(8) Morphological type (see Section \ref{sec:morphtype}.)\\

\tablenotetext{a}{From \citet{IglesiasParamo2002}}
\tablenotetext{b}{Object is near the edge of the observed region, 
and the full extent of the cloud is uncertain}
\tablenotetext{c}{Parent galaxy is uncertain, and the cluster redshift is used.
Derived \Ha\ flux and mass would have large error.}
\tablenotetext{d}{From \citet{Gavazzi2003}}
\tablenotetext{e}{SF blobs are not included.}

\end{table}

\begin{table}[ht]
\caption{Parents of Extended \Ha\ clouds in \A1367}
\label{tab:parents}

\begin{tabular}{|c|c|c|c|c|c|c|c|c|c|}
\hline
parent        
& d
& redshift
& NucAc
& $r$
& $g-r$
& EW(\Ha)
& $M_r$
& log(M$_*$)&type\\
(1)&(2)&(3)&(4)&(5)&(6)&(7)&(8)&(9)&(10)\\
\hline
CGCG~097-073  & 739 &0.02429& HII & 15.5 & 0.38& 111.0 & -19.5 & 9.213 &1\\
CGCG~097-079  & 682 &0.02342& HII & 15.6 & 0.28& 129.0 & -19.4 & 8.963 &1\\
CGCG~097-087  & 467 &0.02218& \nodata & 14.6 & 0.61\tablenotemark{a}& 77.0 & -20.4 & 9.784 &1\\
CGCG~097-087N & 473 &0.02516\tablenotemark{b}&  \nodata&16.5 & 0.43& 19.0 & -18.5 & 9.111 &1\\
CGCG~097-092  & 754 &0.02157& HII & 14.9 & 0.45& 45.17 & -20.1 & 9.544&3\\
\multicolumn{1}{|l|}{2MASX~J11443212} &&&&&&&&&\\
~+2006238
              & 580 &0.02402& HII & 15.1 & 0.82& \nodata & -19.9 & 9.869&2 \\
CGCG~097-093  & 230 &0.01633& HII & 15.1 & 0.26& 9.0 & -19.9 & 9.143&1\\
CGCG~097-122  & 518 &0.01824& HII & 14.1 & 0.64& \nodata & -20.9 & 9.95&1\\
\hline
(BIG)&&&&&&&&&\\
CGCG~097-120  & 102 &0.01871& AGN & 13.0 & 0.73& 4.0 & -22.0 & 10.41&1\\
CGCG~097-114  & 78  &0.02825& HII & 15.4 & 0.36& 36.0 & -19.6 & 9.229&1\\
CGCG~097-125  & 123 &0.02742& HII & 14.3 & 0.76& 23.0 & -20.7 & 10.27&1\\
\multicolumn{1}{|l|}{SDSS~J114501.81} &&&&&&&&&\\
~+194549.4 
              & 166 & 0.02742& BAD & 16.4 & 0.52& \nodata & -18.6 & 9.111&3?\\
\hline
\multicolumn{1}{|l|}{(SDSS~J114513.76} &&&&&&&&&\\
~+194522.1)
              & 244 & 0.02754& HII & 15.3 & 0.69 & 12.0 & -19.7 & 9.678& \nodata\\
\hline
\end{tabular}

(1) Name of the galaxy.\\
(2) Projected distance from the cluster center. Unit is kpc.\\
(3) Redshift. Taken from SDSS DR12\citep{DR12} unless otherwise noted.\\
(4) Nuclear activity from spectroscopy\citep{Gavazzi2011}.\\
(5) r-band magnitude from SDSS DR12\\
(6) g-r color from SDSS DR12\\
(7) Equivalent width of \Ha\ from GOLDMine Database\citep{Goldmine}.\\
(8) r-band absolute magnitude calculated from the photometry in SDSS DR12\citep{DR12}.\\
(9) Log of estimated stellar mass in units of M$_{\sun}$ 
assuming Chabrier IMF.\\
(10) Morphological type (see Section \ref{sec:morphtype}.)
\tablenotetext{a}{From GOLDMine Database\citep{Goldmine}}
\tablenotetext{b}{From \citet{IglesiasParamo2002}}
\end{table}

\begin{table}[ht]
\caption{Candidates of Star-forming blobs in the tail of BIG and near orphan2}
\label{tab:SFRs}
\begin{tabular}{|c|c|c||c|c|c|c||c|c|}
\hline
ID & RA(2000) & DEC(2000) & \Ha\ flux\tablenotemark{a}& B\tablenotemark{b} & R\tablenotemark{b} & i\tablenotemark{b} &  FUV & NUV \\
\hline
B1 & \RAform{11}{44}{45}{3} & \Decform1{+19}{49}{36} & 3$\times 10^{-16}$  & 22.8 & 22.9 & 23.0 &22.43$\pm$0.17 & 22.65$\pm$0.26\\
B2\tablenotemark{c} & \RAform{11}{44}{44}{4} & \Decform1{+19}{49}{39} & 5$\times 10^{-16}$  & 23.0 & 22.6 & 22.4 &22.90$\pm$0.25 & 22.40$\pm$0.24\\
B3 & \RAform{11}{44}{42}{2} & \Decform1{+19}{49}{50} & 2$\times 10^{-16}$  & 23.6 & 23.9 & 24.4 &21.98$\pm$0.18 & 23.34$\pm$0.35\\
B4\tablenotemark{c} & \RAform{11}{44}{39}{4} & \Decform1{+19}{49}{49} & 2$\times 10^{-16}$  & 23.1 & 22.5 & 22.6 &22.36$\pm$0.21 & 22.92$\pm$0.24\\
B5\tablenotemark{c} & \RAform{11}{44}{40}{3} & \Decform1{+19}{50}{46} & 3$\times 10^{-16}$  & 23.1 & 22.6 & 22.8 &22.30$\pm$0.19 & 21.78$\pm$0.18\\
B6 & \RAform{11}{44}{32}{7} & \Decform1{+19}{51}{24} & 3$\times 10^{-16}$  & 23.6 & 23.3 & 23.7 &22.33$\pm$0.16 & 22.67$\pm$0.32\\
\hline
O1 & \RAform{11}{44}{25}{2} & \Decform1{+20}{09}{45} & 6$\times 10^{-16}$ & 20.5 & 20.2  & 19.9 &21.79$\pm$0.15 & 21.49$\pm$0.11\\
\hline
\end{tabular}

\tablenotetext{a}{\Ha\ flux within the isophote of $\times$ 10$^{-18}$ erg s$^{-1}$ cm$^{-2}$ arcsec$^{-2}$.
 Unit is erg s$^{-1}$ cm$^{-2}$. Assuming z=0.0275 for B1--B6, and z=0.0217 for O1.}
\tablenotetext{b}{Measured within \Ha\ isophote, 
and not directly comparable to UV magnitudes.}
\tablenotetext{c}{Blended with red (probably fore/background) objects.}

\end{table}

\begin{table}[ht]
\caption{Coma and A1367 statistics in surveyed region}
\label{tab:stat}
\begin{tabular}{|l|c|c|c|c|c|}
\hline
name & survey area[Mpc$^2$] & member & blue member & EIG parent\tablenotemark{a} & blue parent\tablenotemark{a}\\
\hline
Coma  & 1.2 & 202 & 15 & 12 & 8 \\
A1367 & 1.7 & 120 & 19 & 11 & 6 \\
\hline
\end{tabular}

\tablenotetext{a}{Only $r<17.7$ mag ones are counted.}
\end{table}

\clearpage
\appendix

\section{Mass estimation of the ionized gas clouds}
Using the photometric properties of the clouds given in
Table \ref{tab:clouds}, we estimated their masses as follows.
We assume that the cloud is optically thin,
and the volume of the cloud is calculated as $V = S \times L$,
where $S$ is the projected area given in Table \ref{tab:clouds},
and $L$ is the mean length along the line of sight.
Assuming a roughly cylindrical morphology,
$L$ is approximated by $L = B \times \sqrt{S/(A\times B)}$,
where $A$ and $B$ are the length and the width
of the bounding rectangle, respectively.
The volume is thus calculated as $V = S^{3/2}\sqrt{B/A}$.
The case B recombination coefficient of \Ha\ at $T_e = 10^4$ K is
$\alpha_B \sim 8.7 \times 10^{-14}$ cm$^3$ s$^{-1}$\citep{Osterbrock2006}.
The \Ha\ flux given in Table \ref{tab:clouds} ($f_{H\alpha}$)
is then calculated as
\begin{equation}
f_{H\alpha} =  \frac{h\nu_{H\alpha} \alpha_B n_e^2 V}{4\pi d_L^2},
\end{equation}
where 
$n_e$ is the mean electron density,
$h$ is the Planck constant,
$\nu_{H\alpha}$ is the frequency of \Ha,
and $d_L$ is the luminosity distance to the gas, 100 Mpc.
And the mass ($m$) is calculated as 
\begin{eqnarray}
m
&\sim& 9.7\times 10^{7}M_\sun \times\nonumber\\
&& \sqrt{
f_v}
\sqrt{
\frac{f_{H\alpha}}{10^{-14} {\rm erg s^{-1} cm^{-2}}} 
\left(\frac{S}{100 {\rm kpc}^2}\right)^{3/2} \sqrt{\frac{B}{A}}}
\label{eqn:mass-est}
\end{eqnarray}
where $f_v$ is the volume filling factor of the ionized gas,
and $m_p$ is the proton mass.
The calculated mass of the clouds is given in Table \ref{tab:clouds}.
If the cloud distribution has inclination $i$ against 
the tangential plane, the derived mass should be divided by $\cos i$, 
but this effect is smaller than other uncertainties.

\section{Notes on known EIGs}

\subsection{CGCG~097-073 and 097-079}

\begin{figure}[ht]
\includegraphics[scale=4,bb=0 0 59 61]{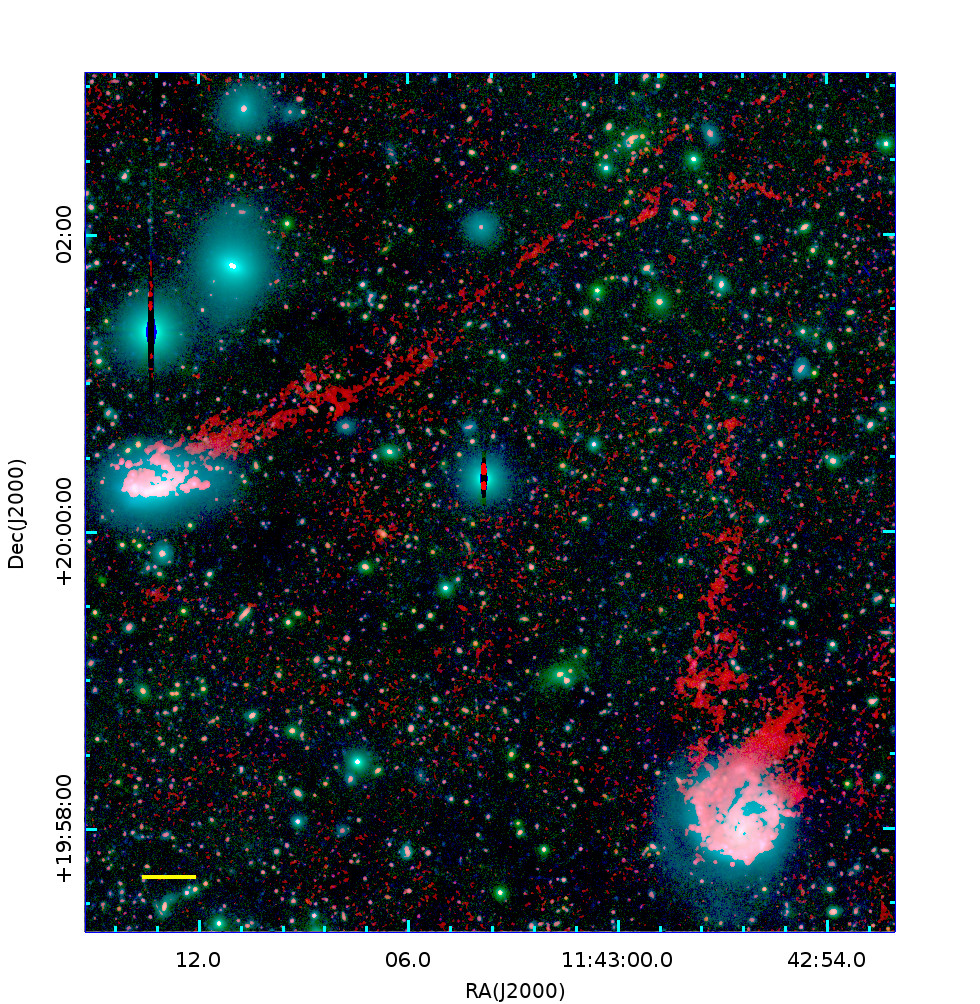}
\includegraphics[scale=4,bb=0 0 59 61]{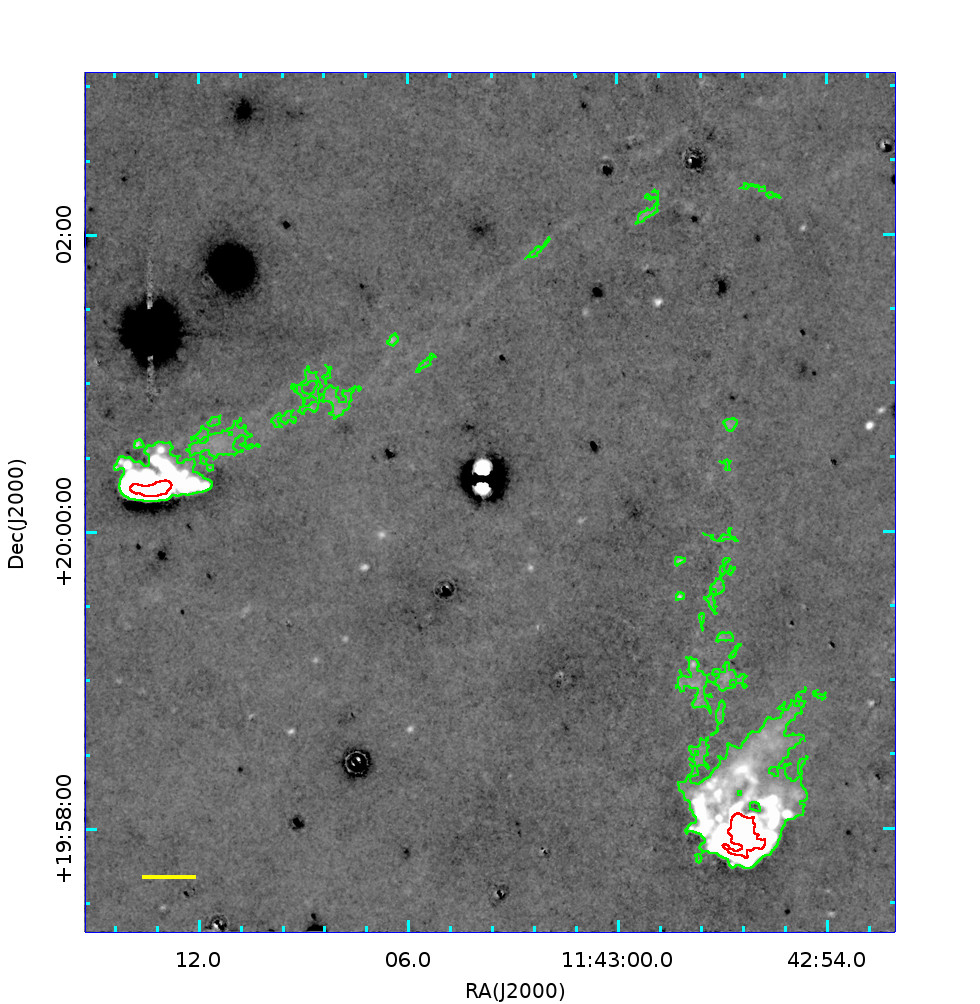}\\
\caption{Same as Figure \ref{fig:097092}, but around 
CGCG~097-079 and CGCG~097-073. 
In the left panel, we can see deep details of the tails.
 Meanwhile, we adopted the shallower isophote 
shown in the right panel for analysis.
As the redshift of the two galaxies is relatively close to the cluster, 
the transmittance at their \Ha\ is the highest in \NB filter, 
and \Ha\ flux is observed as a stronger signal
than those with larger offset in the redshift such as BIG.
For a statistical discussion, we thus used the shallow isophote,
2.5$\times$ 10$^{-18}$ erg s$^{-1}$ cm$^{-2}$ arcsec$^{-2}$,
to keep a comparable depth to the other EIGs.
}
\label{fig:097079}
\end{figure}

The two are well known for their prominent tails 
\citep[e.g.,][]{Gavazzi1987b,Gavazzi1989,Boselli1994,
Gavazzi1995,Gavazzi1998,Gavazzi2001,Scott2010}.
The Subaru images as Figure \ref{fig:097092} is
shown as Figure \ref{fig:097079}.
In \citet{Boselli2014}, we showed a preliminary result
from the same data as this paper.

\subsection{CGCG~097-087 and 097-087N}

CGCG~097-087 (UGC~6697) is also the best-studied galaxy in \A1367
\citep[e.g.,][]{Gavazzi1987b,Gavazzi1989,Boselli1994,Gavazzi1995,Gavazzi1998,Gavazzi2001b,Scott2010}.
\citet{Gavazzi2001b} presented \Ha\ image and spectra of the
galaxy and a part of the tail with a detailed investigation.

\begin{figure}[ht]
\includegraphics[scale=4,bb=0 0 60 54]{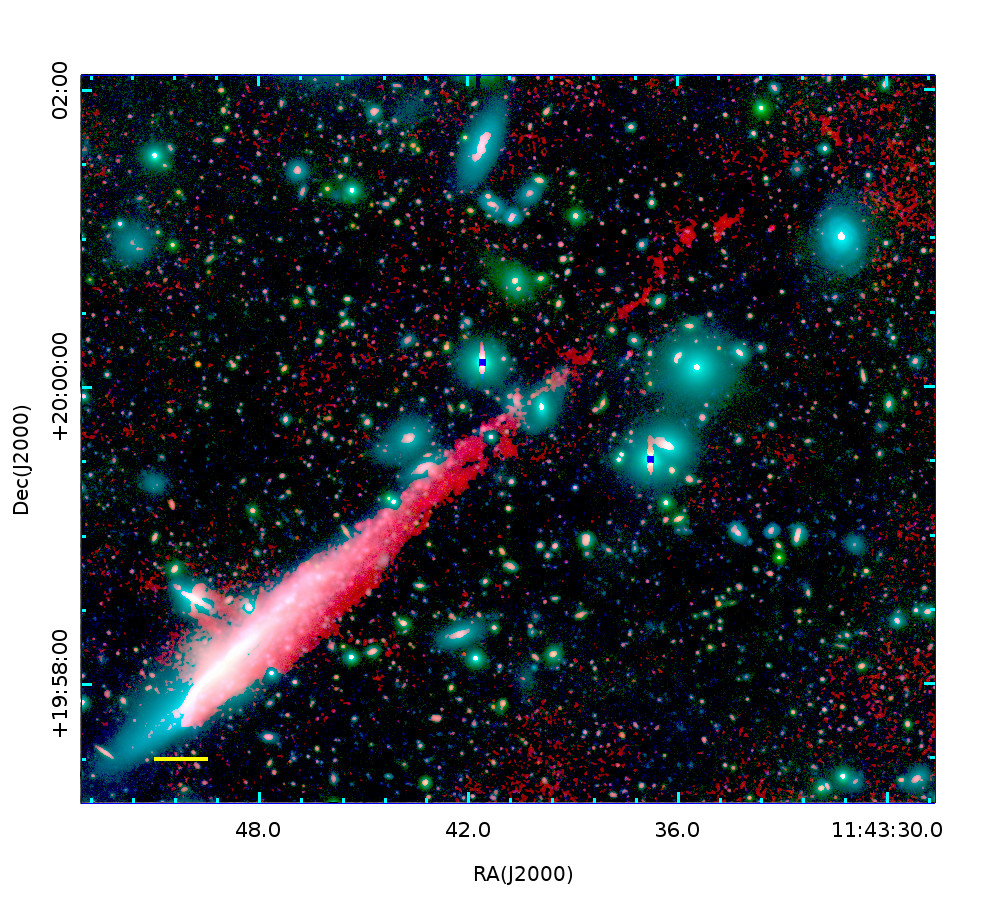}
\includegraphics[scale=4,bb=0 0 60 54]{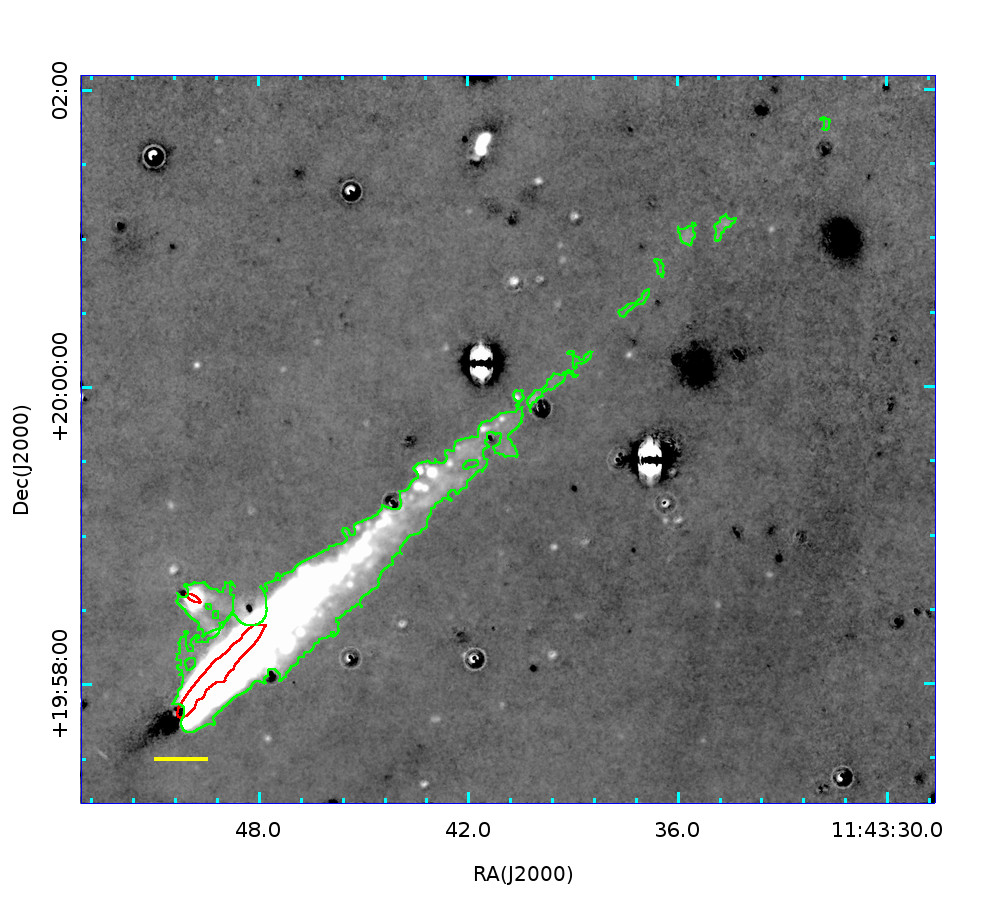}\\
\caption{Same as Figure \ref{fig:097092}, but around CGCG~097-087
and CGCG~097-087N.}
\label{fig:097087}
\end{figure}

In the Subaru \Ha\ data (Figure \ref{fig:097087}, 
we found that the tail of CGCG~097-087 
extends more than $>$150 kpc from the core,
more than twice as long as previously known 
in \Ha\ \citep{Gavazzi1995} 
and in other wavelengths \citep{Sun2005,Scott2010}.

\begin{figure}[htb]
\includegraphics[scale=12,bb=0 0 26 25]{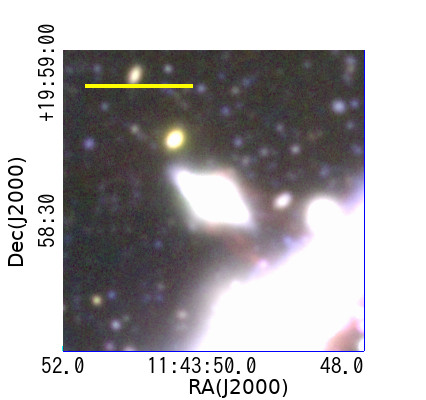}
\caption{Zoomed up images of CGCG~097-087N
\Ha(\NB-R) image before PSF matching.
}
\label{fig:097087N}
\end{figure}

In the Subaru \Ha\ data, 
CGCG~097-087N shows clear twin tails without 
stellar counterpart toward CGCG~097-087 (Figure \ref{fig:097087}).
Zoomed up images before PSF matching are shown as Figure \ref{fig:097087N}.
The existence of the tails that overlap on the 
CGCG~097-087 disk may be a hint to understand 
the complex velocity field of the CGCG~097-087 disk
\citep{Gavazzi1984,Gavazzi2001b}.

CGCG~097-087 shows a quite red color ($g-r$ = 1.81) in DR12 catalog.
We checked the SDSS database to find that
the center of the galaxy suffers heavy dust extinction
and SDSS extracted the region.
The color therefore does not reflect its stellar population.
We therefore used the color from GOLDMine Database \citep{Goldmine}
for the galaxy, which measured SDSS images with 
the method by \citet{Consolandi2016}.
The adopted color of CGCG~097-087 is $g-r = 0.61$.

\subsection{EIG parents in BIG}

For statistical discussion of EIG parents, 
the assignment of the parent galaxy of the complex \Ha\ clouds around BIG
is needed.
From previous studies, it is known that
CGCG~097-125 and CGCG~097-114 show a clear \Ha\ tail 
and were identified as the parents.
CGCG~097-120 ($v_r$ = 5609 km s$^{-1}$; from SDSS DR12) 
has been thought as an accidental overlap,
since the measured recession velocity of the \Ha\ clouds around BIG 
are as high as 8000 -- 8800 km s$^{-1}$ \citep{Cortese2006},
However, a recent spectroscopic observation by MUSE/VLT
revealed that CGCG~097-120 is interacting with 
surrounding \Ha\ gas, and showing a 
smoothly connected distribution of the \Ha\ velocity 
(Consolandi, G. et al. in preparation).
Thus at least these three are parent galaxies of EIGs.

Another possible parent of EIGs around BIG 
is SDSS~J114501.81+194549.4, which shows clear poststarburst feature
from SDSS DR12 spectrum at $v_r = 8220$ km s$^{-1}$.
The EIG just south of the galaxy may be connected with the galaxy.
The galaxy is thus counted as an EIG parent in Table \ref{tab:stat}.

Yet another possible parent of EIG around BIG is SDSS~J114513.76+194522.1
($v_r = 8256$ km s$^{-1}$)
about 150 kpc east of the BIG complex.
The galaxy shows a sign of stripping toward west, 
though no extended \Ha\ feature is seen around it.
Moreover the South clouds of BIG \citep{Cortese2006} 
show a sign that they come from far east of BIG.
These are, however, currently weak evidence.
Thus we did not count the galaxy in Table \ref{tab:stat}
and did not plot in Figures \ref{fig:CMD} and \ref{fig:vd},
but showed it in Table \ref{tab:parents} with parentheses 
as a reference.

Other spectroscopically confirmed dwarfs and knots 
by \citet{Sakai2002} and \citet{Cortese2006} 
are fainter than the magnitude limit of the catalog ($r<17.7$)
used in the statistical analysis of this study.
The previous studies 
\citep{Sakai2002,Gavazzi2003,Cortese2006} discussed 
the possibility that the dwarfs could have been formed in the stripped 
gas from giant galaxies.
The dwarfs would not be parent galaxies of EIGs but 
rather children.
We therefore do not count them as parent galaxies of EIGs.

\clearpage

\end{document}